\documentclass[aps,twocolumn,longbibliography] {revtex4-2}
\usepackage{amsmath,amssymb,graphicx,amsfonts}

\usepackage{times}
\usepackage[colorlinks=true,linkcolor=blue,urlcolor=blue,citecolor=blue,breaklinks=true]{hyperref}

\usepackage{comment}

\usepackage[utf8]{inputenc}

\def\da{\dagger}

\def\eq#1{Eq.~(\ref{#1})}
\def\da{\dagger}

\date\today
\begin{document}

\title{Relaxation dynamics of a mobile impurity injected into a one-dimensional Bose gas}

\author{Saptarshi Majumdar}
\author{Aleksandra Petkovi\'{c}}
\affiliation{Univ Toulouse, CNRS, Laboratoire de Physique Théorique, Toulouse, France}

\begin{abstract}
The nonequilibrium dynamics of an impurity immersed with a finite velocity into a one-dimensional system of weakly interacting bosons is studied within the framework of the time-dependent Gross-Pitaevskii equation. We uncover and characterize different regimes of relaxation dynamics. 
We find that the final impurity velocity remains constant in a large interval of sufficiently big and realistic initial velocities. The underlying physical mechanism is the emission of  the dispersive density shock waves that carry away the excess of the initial impurity momentum, while locally the system remains in the same stationary state. In contrast, a heavy impurity with the same coupling constant relaxes differently and the regime of constant final velocity disappears. Furthermore, a fast heavy impurity exhibits damped velocity oscillations in time before reaching a stationary state. This process is accompanied by the oscillations of the local depletion of the boson density around the impurity, until their positions coincide and they continue the motion together.
Decreasing the impurity-boson coupling or increasing the strength of repulsion between bosons,  the oscillations get amplified.  In the case of a heavy impurity with the mass bigger than the critical one, the ground state energy as a function of momentum exhibits cusps and metastable branches. We show that they manifest themselves by a soliton emission, a considerable slowing down of the relaxation, and a change of the impurity direction of motion with respect to the initial one. 
\end{abstract}

\maketitle

\section{Introduction}

Understanding the dynamics of a foreign particle (an impurity) through a many-body environment is a fundamental question, relevant for a large class of systems. Particularly interesting environments are one-dimensional quantum liquids where some spectacular phenomena take place \cite{zvonarev2007spin,AnnalsKamenev,peotta2013quantum,PhysRevLett.113.070601,PhysRevA.92.023623,Grusdt_2017,LewensteinBrownien,PhysicsReports2023,Gamayun2024}. One such example are the Bloch oscillations of an impurity subjected to a small external force in the absence of a lattice, where the impurity oscillates around a fixed position in space while the supercurrents carry the momentum pumped into the system \cite{gangardt2009bloch,Meinert945}. This phenomenon is closely related to the peculiar property of
one-dimensional many-body systems in the presence of an impurity, where
the lowest energy state for a given momentum is a periodic
function of momentum \cite{lamacraft2009dispersion,kamenev2009dynamics,AnnalsKamenev}.  

Another important problem is the dynamics of fast impurities. It is well known that the motion of fast obstacles causes some intriguing phenomena, such as the sonic boom provoked by the supersonic airplane motion and the Cherenkov radiation emitted by a superluminal charged particle \cite{Landau}. Related phenomena take also place in quantum mechanical systems, for example during the motion of an obstacle \cite{landau+49,Hakim,PhysRevA.66.013610,astrakharchik2004motion,PRLCarusotto2006,Cherny2012} or a finite-mass impurity \cite{QFlutterNature,PhysRevE.90.032132,knap2014quantum,robinson2016motion,QuenchZvonarev,atoms10010003,Will_2023,YangGaudinFlutter} through a superfluid system, with a velocity exceeding the critical one.
In the case of a supersonic impurity moving through a one-dimensional system of strongly interacting bosons,  the impurity velocity undergoes slowly-decaying oscillations in time \cite{QFlutterNature,knap2014quantum,YangGaudinFlutter}.

The present work studies the real-time dynamics of a finite-mass impurity injected with a finite velocity into a system of weakly interacting bosons. The bosons are assumed to be in their ground state at the moment of the impurity injection. We consider the evolution from this far-from-equilibrium initial state as a function of interaction strengths, the initial impurity velocity and its mass.
Our goal is to uncover and characterize different dynamical regimes,  to understand corresponding mechanisms and characteristic time scales of relaxation, as well as the final stationary state in different regions of parameters. We solve the time-dependent mean-field equation of motion for bosons in the reference frame co-moving with the impurity. This approach allows us to study a wide range of impurity-boson interaction strength, initial velocity and impurity mass. We unveil a very rich relaxation dynamics of the system. 

Cold atomic gases provide an ideal playground for the study of the nonequilibrium impurity dynamics, due to possibility to control and manipulate the quantum entities on
the atomic size, developed advanced spectroscopic methods and high-resolution imaging techniques, as well as the possibility to fine-tune interactions using the Feshbach resonances \cite{grusdt2025impurities}.
Experimental realizations of one-dimensional systems of bosons with impurities thus provide a promising testing bed for our predictions \cite{palzer2009quantum,2012quantum,fukuhara2013quantum,Meinert945}.

The paper is organized as follows. Sec.~\ref{sec:model} introduces the model. In Sec.~\ref{sec:mean-field}, we study the mean-field equation of motion and its variations. Namely, we point out some shortcomings of a widely used equation of motion and propose another one. For both equations, we review the properties of the stationary states.
Sec.~\ref{sec:quench} investigates the time-evolution after the impurity-boson interaction is quenched on, by providing numerical solutions of the time-dependent equations of motion. Initially, the impurity has a finite velocity, while the bosons are in their ground state. We present and characterize a panorama of different regimes. The main results are as follows: (i) We show that in a large interval of sufficiently big initial impurity velocities, the final impurity velocity remains almost constant while the remaining momentum is carried by the dispersive density shock waves. The dynamics of heavy impurities is substantially different, and this regime disappears.
(ii) We demonstrate that a hydrodynamic theory reveals temporal velocity oscillations of fast heavy impurities, and show that the oscillations are amplified at weak impurity-boson coupling. This phenomenon was predicted for strongly coupled fast impurities with the mass similar to the mass of the bosons, immersed into a system of strongly interacting bosons \cite{QFlutterNature,knap2014quantum}. It was stressed that this phenomenon can not to be captured by a hydrodynamic approach \cite{QFlutterNature}.  (iii) A very heavy impurity, with mass bigger than the critical one, relaxes by emitting both solitons and density waves. In the close vicinity of the termination points of the metastable branches, the relaxation slows down significantly and the impurity changes the initial direction of motion. The paper concludes with a summary of main findings and their discussion presented in Sec.~\ref{sec:conclusions}.

\section{Model\label{sec:model}}

We study a system of one-dimensional bosons with contact repulsion in the presence of a single mobile impurity at zero temperature. The system is modelled by the Hamiltonian
\begin{align}
\hat{H}=\hat{H}_b+\frac{\hat{P}^2}{2M}+G \hat{\Psi}^\da(\hat{X})\hat{\Psi}(\hat{X}).
\label{eq:H}
\end{align}
Here $\hat H_b$ describes the Bose gas and reads as 
\begin{align}
\hat{H}_b=\int \mathrm{d}x\left[-\hat\Psi^\da(x)\dfrac{\hbar^2\partial_x^2}{2m}\hat\Psi(x)+\frac{g}{2}\hat\Psi^\da(x)\hat\Psi^\da(x)\hat\Psi(x)\hat\Psi(x)\right].
\end{align}
The bosonic single-particle operators  $\hat\Psi^\da(x)$ and $\hat\Psi(x)$ satisfy the commutation relation $[ \hat\Psi(x) , \hat\Psi^\da(x')]=\delta(x-x')$. The repulsion between the bosons has the strength $g$ which enters the dimensionless parameter $\gamma=m g/\hbar^2 n_0$. Here $m$ denotes the mass of bosons and $n_0$ is the mean boson density. The impurity of mass $M$ couples locally to the density of bosons, with a coupling constant $G$.  We assume that the impurity-boson interaction is repulsive, $G>0$. The impurity momentum and position operators are denoted by $\hat{P}$ and $\hat X$, respectively. 

We apply the Lee-Low-Pines transformation \cite{LeeLowPines} that acts as $\hat{\mathcal{H}}=\hat{U}^\da \hat{H}\hat{U}$ where $\hat{U}=e^{-i \hat{X} \hat{p}_b/\hbar}$. Here, the momentum of the Bose gas is denoted by $\hat{p}_b=-i\hbar\int \mathrm{d}x  \hat\Psi^\da(x)\partial_x \hat\Psi(x)$.
Then, the impurity momentum is transformed as $\hat{U}^\da \hat{P}\hat{U}=\hat{P}-\hat{p}_b$. 
For bosons, this transformation is actually a translation by the impurity position $\hat{X}$, $\hat{U}^\da \hat{\Psi}(x)\hat{U}=\hat{\Psi}(x-\hat{X})$, and in the new referent system the impurity is situated at the origin. As a result, the Hamiltonian $\hat{\mathcal{H}}$ does not depend on $\hat{X}$, and $\hat{P}$ is conserved in time and can be replaced by a number $p$.
This signals the conservation of the total system momentum, which in the reference co-moving with the impurity is given by
$
\hat{U}^\da  (\hat{P}+\hat{p}_b) \hat{U}=\hat{P}.
$
The final Hamiltonian takes the form 
\begin{align}\label{eq: HLee-Low}
\hat{\mathcal{H}}=\hat{H}_b+\frac{(p-\hat{p}_b)^2}{2M}+G \hat{\Psi}^\da(0)\hat{\Psi}(0),
\end{align}
where $p$ is the total momentum of the system. 
We consider the following problem. The impurity-boson coupling is switched on at $t=0$. At that moment, the impurity has an initial momentum $p=M V_0$, while the bosons are in their ground state and have zero momentum. We are interested in the evolution of the system from this initial state.

\section{Stationary state properties\label{sec:mean-field}}

\subsection{Mean-field equation}

It is convenient to work in the Heisenberg picture, where the single-particle operator $\hat{\Psi}$ depends on time. Its equation of motion is given by $i\hbar\partial_t\hat\Psi=[\hat\Psi,\hat{\mathcal{H}}]$ where all the operators are considered in the Heisenberg picture. It reads as
\begin{align}
i\hbar \partial_t{\hat{\Psi}(x,t)}=&\Bigg[ -\frac{\hbar^2}{2m}\partial_x^2+g |\hat{\Psi}(x,t)|^2+G\delta(x)\Bigg] \hat{\Psi}(x,t)
\notag\\&+\frac{i\hbar}{2}\Big\{\hat{V}(t) [\partial_x \hat\Psi(x,t)]+[\partial_x \hat\Psi(x,t)]\hat{V}(t) \big\}.
\label{eq:motion}
\end{align}
Having in mind that the total momentum is conserved, and that one part of the initial impurity momentum is channeled into the bath of bosons, we introduce the impurity velocity $\hat{V}(t)=(p-\hat{p}_b(t))/M$. We are interested in weakly-interacting bosons, $\gamma\ll 1$. The field operator  can be expanded as \cite{pitaevskii_bose-einstein_2003,sykes_drag_2009,CasimirNewJPhys}
	$
		\hat\Psi(x,t)=\Psi_0(x,t)+\gamma^{1/4}\hat\Psi_1(x,t)+\ldots, \label{eq:wf_exp}
	$
where the field $\Psi_0(x,t)$ describes the condensate wave function in the absence of  fluctuations, while higher order contributions account for the effects of quantum and thermal fluctuations.  
At the mean-field level, the equation of motion reads as \cite{atoms10010003,Will_2023}
\begin{align}
i\hbar \partial_t{{\Psi_0}(x,t)}=\Bigg[&-\frac{\hbar^2}{2}\left( \frac{1}{m}+\frac{1}{M}\right)\partial_x^2+g |{\Psi_0}(x,t)|^2\notag\\&+G\delta(x)+i\hbar V(t)\partial_x\Bigg] {\Psi_0}(x,t).
\label{eq:mean-field1}
\end{align}
Here, the impurity velocity takes the form 
\begin{align}\label{eq:Vimp}
V(t)=\frac{p}{M}+i\frac{\hbar}{M}\int \mathrm{d}x  \Psi_0^*(x,t)\partial_x \Psi_0(x,t).
\end{align}
Eq.~(\ref{eq:mean-field1})  can also be obtained by generalizing the approach of Ref.~\cite{VolosnievPolaron} to the case of a nonzero total system momentum \cite{PhysRevA.100.013619}.

Additionally, we will also consider the  equation
\begin{align}
i\hbar \partial_t{{\Psi_0}(x,t)}=\Bigg[&-\frac{\hbar^2}{2m}\partial_x^2+g |{\Psi_0}(x,t)|^2+G\delta(x)\notag\\&+i\hbar V(t)\partial_x\Bigg] {\Psi_0}(x,t).
\label{eq:mean-field2}
\end{align}
Note that the term $-\hbar^2\partial_x^2\Psi_0/(2M)$ in Eq.~(\ref{eq:mean-field1}) is obtained from Eq.~(\ref{eq:motion}) by first rewriting the second line as $i\hbar\hat{V}(\partial_x \hat\Psi)-\hbar^2\partial_x^2\hat\Psi/(2M)$ and then replacing $\hat\Psi$ by $\Psi_0$.  However, if the replacement of the single-particle field operator by the wavefunction is done on the preferred symmetric form $\hat{V}(\partial_x \hat\Psi)+(\partial_x \hat\Psi)\hat{V}$ in Eq.~(\ref{eq:motion}), it leads to \eq{eq:mean-field2}.
The two equations, Eq.~(\ref{eq:mean-field1}) and (\ref{eq:mean-field2}), differ in the first term on the right-hand side. Furthermore, the sound velocity in the laboratory frame that follows from Eq.~(\ref{eq:mean-field2}) is correct and reads as $v=\sqrt{g n_0/m}$ \cite{pitaevskii_bose-einstein_2003}, while Eq.~(\ref{eq:mean-field1}) gives the modified sound velocity $v\sqrt{1+m/M}$ even at vanishingly small $G$. In Sec.~\ref{sec:stationary2}, we will further motivate the analysis of Eq.~(\ref{eq:mean-field2}). 

Note that decreasing the impurity mass, the corrections originating from quantum fluctuations become more relevant. The mean-field approach thus does not hold for a light weakly coupled impurity. However, the effective impurity mass increases with the impurity coupling constant, and the approach is valid for a strongly coupled light impurity.

\subsection{Stationary solution of \eq{eq:mean-field1} \label{sec:stationary1}}

In this section, we consider the stationary solution of \eq{eq:mean-field1} with periodic boundary conditions. We review its properties needed for Sec.~\ref{sec:quench}, where we study the dynamics after a quench. 
The stationary solution can be written as   \cite{pitaevskii_bose-einstein_2003}
$
\Psi_0(x,t)=\Psi_0(x)e^{-i \mu t/\hbar},
$
where $\mu$ denotes the chemical potential of bosons. We assume that the impurity velocity has reached its final constant value $V_f$. Under these assumptions, Eq.~(\ref{eq:mean-field2}) describes the problem of an obstacle modelled by a delta-like potential moving with a constant velocity $V_f$. It has been solved by \citet{Hakim}. The generalisation of this solution for Eq.~(\ref{eq:mean-field1}) is straightforward. 

In the absence of the impurity potential, the solution of Eq.~(\ref{eq:mean-field1}) is given by a soliton moving with the velocity $V_f$ in the laboratory frame \cite{tsuzuki_nonlinear_1971}. In the presence of the impurity,  one looks for the solutions on the left and right-hand side of the impurity in the form of solitons by introducing a constant shift of its position and its phase. These parameters are evaluated by imposing the continuity of the wavefunction and a jump of its first spatial derivative at origin.
The latter condition is obtained by integrating \eq{eq:mean-field1} over $x$, and reads as
$\partial_x\Psi_0(0^+)-\partial_x\Psi_0(0^-)={2mG \Psi_0(0)}/{\hbar^2(1+m/M)}$.
The amplitude of the jump is proportional to the impurity-boson coupling $G$ that can be expressed as $G=-\hbar^2 \left({M}^{-1}+{m}^{-1}\right)/a_{ib}$, where $a_{ib}$ denotes the one-dimensional impurity-boson scattering length \cite{olshanii_1998,Peano_2005,RevModPhys.80.885}. 
Finally, the stationary solution of Eq.~(\ref{eq:mean-field1}) takes the form
\begin{align}
\Psi_0(x)=&\sqrt{\frac{\mu}{g}}\exp{\left[i\; \text{sgn}(x) \arctan\left(\frac{b \tanh{x_0}}{a}\right)\right]}\notag\\ &\times\left[a-i \;b \;\text{sgn}(x)\tanh{\left(\frac{b|x|}{\xi\sqrt{1+m/M}} +x_0\right)}\right]
\label{eq:MFsolution}
\end{align}
in the thermodynamic limit.
Here the healing length is $\xi=\hbar/\sqrt{m \mu}$, and the sound velocity is $v=\sqrt{\mu/m}$. We defined $a={V_f}/{v\sqrt{1+m/M}}$ and $b=\sqrt{1-a^2}$.
From the condition on the jump of the first spatial derivative of the wavefunction at origin, it follows that the shift parameter $x_0$ satisfies the condition
\begin{align}
\frac{G}{\hbar v}\frac{1}{\sqrt{1+m/M}}(a^2+b^2 \tanh^2{x_0})=b^3 \tanh{x_0}\; \textrm{sech}^2{x_0}.
\label{eq:xo}
\end{align}
Note that \eq{eq:xo} admits solutions only for the impurity velocities smaller than the critical velocity, $|V_f|\leq v_c$.  At higher velocities no stationary state exists, as will be discussed in more detail in Sec.~\ref{sec:quench} where the time-dependent equation of motion for the condensate wave function is solved. 
The critical velocity satisfies \cite{Hakim}
\begin{align}\label{eq:vc}
	\frac{G}{\hbar v}\frac{1}{\sqrt{1+m/M}}=\frac{\sqrt{1-20
			\tilde{v}_c^2-8 \tilde{v}_c^4+\left(1+8 \tilde{v}_c^2\right)^{3/2}}}{2 \sqrt{2}\tilde{v}_c}.
\end{align}
Here we have introduced $\tilde{v}_c=v_c/v \sqrt{1+m/M}$.  From \eq{eq:vc}, it follows that $v_c$ is a decreasing function of the impurity mass as well as the dimensionless impurity strength $\tilde{G}=G/\hbar v$.

Apart from modifying the boson density, the impurity also affects the phase of the condensate. It causes the phase drop $\theta$ across its position. In the thermodynamic limit, $\theta$ reads as 
\begin{align}
\theta=&2 \arctan \left(\frac{b}{a} \right) -2\arctan\left(\frac{b }{a}\tanh{x_0}\right).
\label{eq:theta}
\end{align} 
The periodic boundary conditions thus impose an additional $\theta x/L$ contribution in the phase of the condensate wave function (\ref{eq:MFsolution}) \cite{AnnalsKamenev,PhysRevResearch.2.033142}. Here the system length is denoted by $L$. Note that this term contributes to the system momentum (\ref{eq:momentum}), but does not contribute to the energy (\ref{eq:PolaronEnergy}).  It actually gives the momentum carried by the supercurrents.

The chemical potential is determined by the equation
$\int_{-L/2}^{L/2} \mathrm{d} x |\Psi_0(x)|^2=n_0 L$,
and takes the form
$\mu=g n_0+2 b \hbar  ({1-\tanh{x_0}}) \sqrt{{g n_0 (m+M)}/{(m M)}}/L$.
Here $b$ and $x_0$ are evaluated by replacing $\mu$ by its leading order term $\mu_0=gn_0$. From now on, this will be the case for all the parameters if not stated differently, since we are not interested in finite size corrections. We have provided the first finite size correction in $\mu$, since it is needed for the subsequent evaluation of the energy.

We are now ready to obtain the relation between the initial impurity velocity $V_0$  and its final velocity $V_f$ using Eq.~(\ref{eq:Vimp}). It reads as \cite{AnnalsKamenev}
\begin{align}
MV_0=&M{V}_f - {2\hbar n_0}ab \left(1 - \tanh{x_0} \right) +\hbar n_0 \theta.
\label{eq:momentum}
\end{align}
We stress that $V_f$ does not correspond to the final impurity velocity for the dynamics after a quench, as will be shown in Sec.~\ref{sec:crossover}. 
Next, we define the polaron energy as a difference of the ground state energy of the Hamiltonian (\ref{eq: HLee-Low}) at a given total momentum $p=M V_0$ and its ground state energy at zero momentum in the absence of the impurity. We evaluate it to be \cite{AnnalsKamenev}
\begin{widetext}
\begin{align}
\frac{E_{p}}{g n_0}=-\frac{2}{3}b^3\frac{\sqrt{1+m/M} }{\sqrt{\gamma }}\left[\tanh
^3(x_0)-1\right]+\frac{1}{3} b^3
\frac{\sqrt{1+m/M} }{\sqrt{\gamma }} \left[\tanh ^3(x_0)-3 \tanh
(x_0)+2\right]+\frac{MV_f^2}{2g n_0},
\label{eq:PolaronEnergy}
\end{align}
\end{widetext}
where $V_f$ satisfies Eq.~(\ref{eq:momentum}) and $x_0$ obeys \eq{eq:xo}. 
The Hellmann–Feynman theorem reads as $\mathrm{d} E_{p}/ \mathrm{d}p=\langle\mathrm{d} \mathcal{\hat{H}}/ \mathrm{d}p \rangle_p$, where the expectation value is evaluated in the ground state of the system with momentum $p$. Thus, the impurity velocity is $V_f= \mathrm{d}  E_{p}/\mathrm{d}p$, and
the critical velocity $v_c$ is the maximal slope of the energy dispersion as a function of system momentum $E_p(p)$. 
At zero momentum, Eq.~(\ref{eq:PolaronEnergy}) simplifies to the binding energy of a finite-mass impurity obtained in Refs.~\cite{VolosnievPolaron,Pastukhov2019}, as well as an infinitely heavy impurity obtained in Ref.~\cite{PetkovicPRAL}. The expression for the energy reported in Ref.~\cite{Will_2023} differs from \eq{eq:PolaronEnergy} and does not satisfy the condition $V_f= \mathrm{d}  E_{p}/\mathrm{d}p$. The binding energy has been studied numerically in Refs.~\cite{PhysRevA.95.023619,Grusdt_2017}. Numerically exact results for the polaron energy for a small system size are given in Ref.~\cite{condmat7010015}.
\begin{figure}
\centering
    \includegraphics[width= \columnwidth]{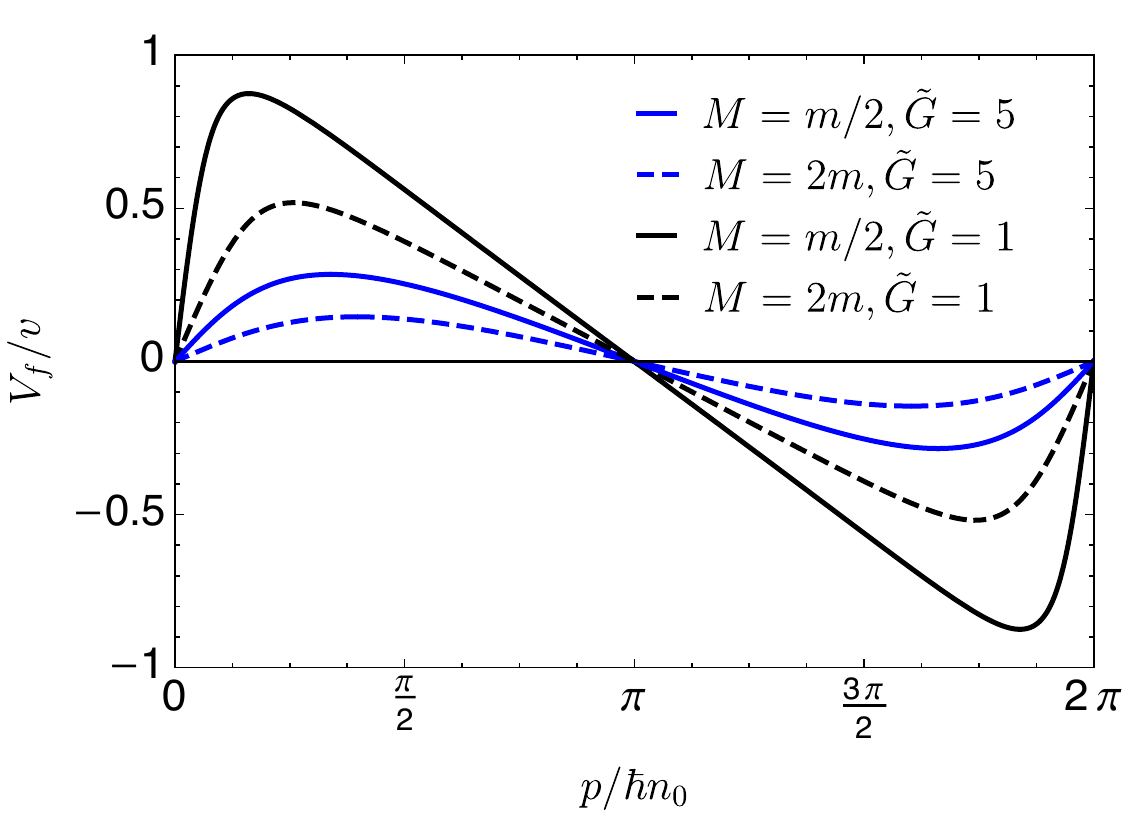}
       \caption{Final impurity velocity as a function of the initial impurity momentum given by \eq{eq:momentum} for $M=m/2$ (solid line) and $M=2m$ (dashed line) for two different values of the dimensionless impurity coupling constant $\tilde{G}=G/\hbar v$. Here $\gamma = 0.1$.}
    \label{fig:vf}
\end{figure}

In Eq.~(\ref{eq:momentum}) the impurity velocity satisfies $|V_f|\leq v_c$  and the momentum is $|p|\leq \pi \hbar n_0$. However, if the momentum of each particle is increased for a minimal value of  $2\pi\hbar/L$, the new wavefunction is obtained by multiplying $(\ref{eq:MFsolution})$ 
by $\exp{(i 2\pi x/L)}$  and the total momentum (\ref{eq:momentum}) is increased by $2\pi\hbar n_0$ without modifying the energy (\ref{eq:PolaronEnergy}).
Thus the energy is a periodic function of  the momentum with a period $2\pi\hbar n_0$ \cite{lamacraft2009dispersion,kamenev2009dynamics,AnnalsKamenev}. Note that at zero temperature dynamic correlation functions exhibit power-law singularities at the polaron spectrum  that mark the edge below which there are no many-body excitations \cite{zvonarev2007spin,matveev2008spectral,kamenev2009dynamics}.

Final impurity velocity $V_f$ as a function of its initial momentum for different impurity masses and impurity coupling constants is shown in Fig.~\ref{fig:vf} for one period.
Note that $V_f$ vanishes at $p=0$, $p=\pi \hbar n_0$ and $p=2\pi \hbar n_0$. Besides, as the impurity coupling constant grows, $V_f$ decreases at each $p$ and at each $m/M$. 
Moreover, for a given initial impurity momentum, $V_f$ is a decreasing function of $M/m$, while for a given initial impurity velocity it is a non-monotonic function of $M/m$.

\begin{figure}
\centering
    \includegraphics[width= 0.98\columnwidth]{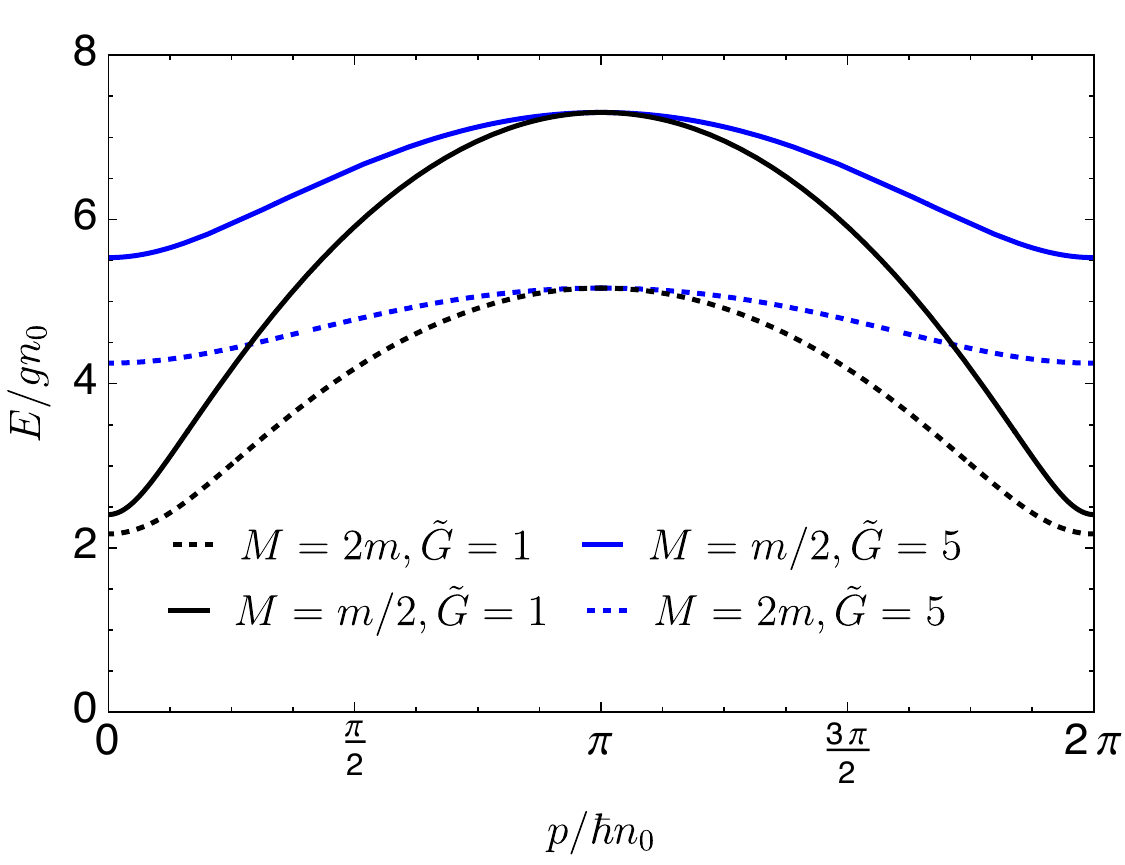}
    \caption{Polaron energy  (\ref{eq:PolaronEnergy}) as a function of  the initial impurity momentum for $M=m/2$ (solid line) and $M=2m$ (dashed line) for two different values of the dimensionless impurity coupling constant $\tilde{G}=G/\hbar v$. Here $\gamma = 0.1$.}
    \label{fig:energy}
\end{figure}

Note that there are two physical solutions of Eq.~(\ref{eq:xo}) with ${x_0}$ being real.
They provide two stationary solutions for a given $V_f$, and they smoothly coincide at $V_f=v_c$. One solution is responsible for one part of the $V_f>0$ curve in Fig.~\ref{fig:vf} where the final impurity velocity increases from zero and reaches its maximal value $v_c$. We refer to it as the first solution. The second solution is realised where $V_f$ decreases from $v_c$ to $-v_c$ as $p$ is increased, and finally the first solution appears again for $V_f$ increasing from $-v_c$ to zero. The polaron energy (\ref{eq:PolaronEnergy}) as a function of momentum is shown in Fig.~\ref{fig:energy}. Note that the energy dispersion at the crossover from the first to the second solution has nonperturbative character in the impurity-boson coupling constant \cite{PRBBose-Fermi}.

In the limit of vanishing impurity strength, $x_0$ of the first solution becomes infinitely big and describes the nonperturbed condensate density $|\Psi_0(x)|=1$. In contrast, $x_0$ of the second solution tends to $0$  and describes a soliton that moves together with the impurity with a maximal density depletion fixed at the impurity position.  
The phase drop (\ref{eq:theta}) and the density depletion at impurity position are increasing functions of the total system momentum $p$ for $0\leq p\leq \pi \hbar n_0$, and reach their maximal value $\pi$ and $n_0$, respectively. Thus at  $p= \pi \hbar n_0$ the second stationary solution is a dark soliton.

\begin{figure*}
	\includegraphics[width= 2\columnwidth]{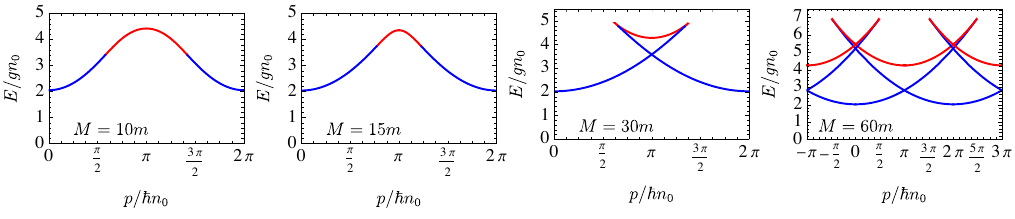}  
	\caption{Energy dispersion (\ref{eq:PolaronEnergy}) for $\tilde{G}=1$ and $\gamma=0.1$ for different impurity masses. The blue and red lines denote the first and the second stationary states, respectively.}
	\label{fig:nonanalytic}
\end{figure*}

Next we consider a heavy impurity. Fig.~\ref{fig:nonanalytic} shows how the polaron energy dispersion changes by increasing the impurity mass while keeping other parameters fixed. The color code is the following: the blue curve corresponds to the first and the red to the second stationary state. 
We see that as $M/m$ increases, the region of the second state decreases and the
blue curves approach each other. At a critical mass, they cross.
Thus, the energy of the ground state becomes nonanalytic function of momentum and develops cusps at $p=(2k+1)\pi \hbar n_0$ with $k$ being an integer, signalling a first order transition \cite{lamacraft2009dispersion,PhysRevLett.108.207001}. The latter is not washed out by quantum fluctuations in the case of a sufficiently weak impurity coupling constant \cite{lamacraft2009dispersion}. We see that the first state is the ground state, while the second one has higher energy and is metastable. The critical mass $M_c$ is determined by the condition $M_c V_0=\pi \hbar n_0$, where $V_0$ is  given by \eq{eq:momentum} and evaluated at $M=M_c$, $V_f=v_c$ and $x_0$ corresponding to the first stationary solution.
If we further increase the impurity mass,  the energy curves enter into neighboring momentum periods and thus we show larger interval of momenta for $M=60m$. The energy landscape becomes very rich with many metastable branches. Note that Ref.~\cite{Hakim} points out that the second stationary state is unstable in the case of $M=\infty$, in agreement with Fig.~\ref{fig:nonanalytic}.

For $M>M_c$, the critical velocity is not anymore given by \eq{eq:vc} once the quantum fluctuations are taken into account. \eq{eq:vc} gives the final impurity velocity at the end of the blue curve where it merges with the red one, which is not the true ground state anymore. Quantum fluctuations allow for tunnelling through the barrier that separates the metastable and the ground states, and system dissipates energy while relaxing into the ground state. Thus the critical velocity, below which an impurity moves without a friction at zero temperature, is $V_f$ given by \eq{eq:momentum} evaluated at $M V_0=\pi \hbar n_0$ \cite{PhysRevLett.108.207001}. 
Note that the critical velocity decreases as $M$ increases and becomes zero in the limit of infinitely heavy impurity, see Fig.~\ref{fig:nonanalytic}. These findings are in agreement with Refs.~\cite{astrakharchik2004motion,Cherny2012} that report a nonzero friction force acting on an infinitely heavy impurity (obstacle) for any finite velocity once the quantum fluctuations are taken into account. 
We will discuss manifestations of the metastable branches in the impurity relaxation dynamics in the following section.

\subsection{Stationary solution of \eq{eq:mean-field2}\label{sec:stationary2}}

\begin{figure}
\centering
    \includegraphics[width= \columnwidth]{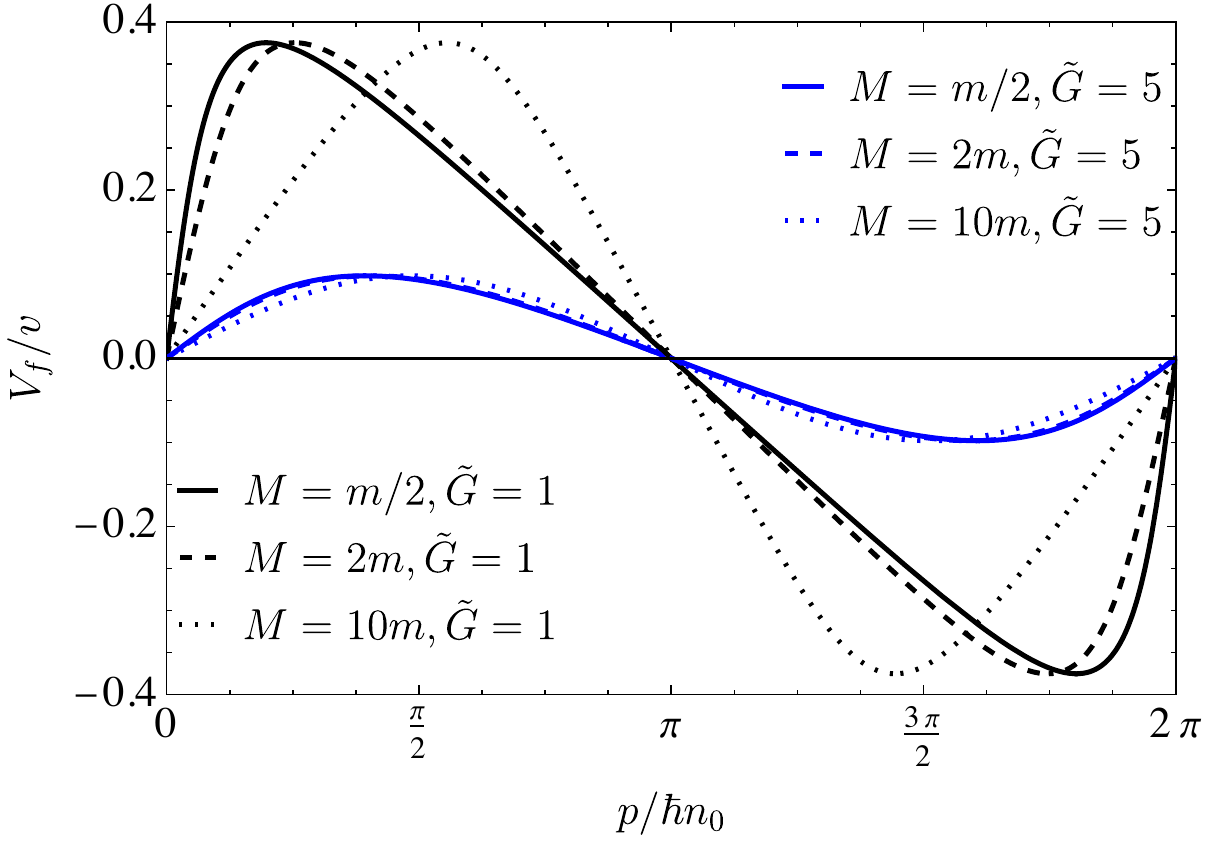}
    \caption{Final impurity velocity for the solution of \eq{eq:mean-field2} as a function of the initial impurity momentum for $M=m/2$ (solid line), $M=2m$ (dashed line) and $M=10m$ (dotted line) for two different values of the dimensionless impurity coupling $\tilde{G}$. Here $\gamma = 0.1$.}
    \label{fig:vf2}
\end{figure}

\begin{figure}
\centering
    \includegraphics[width=\columnwidth]{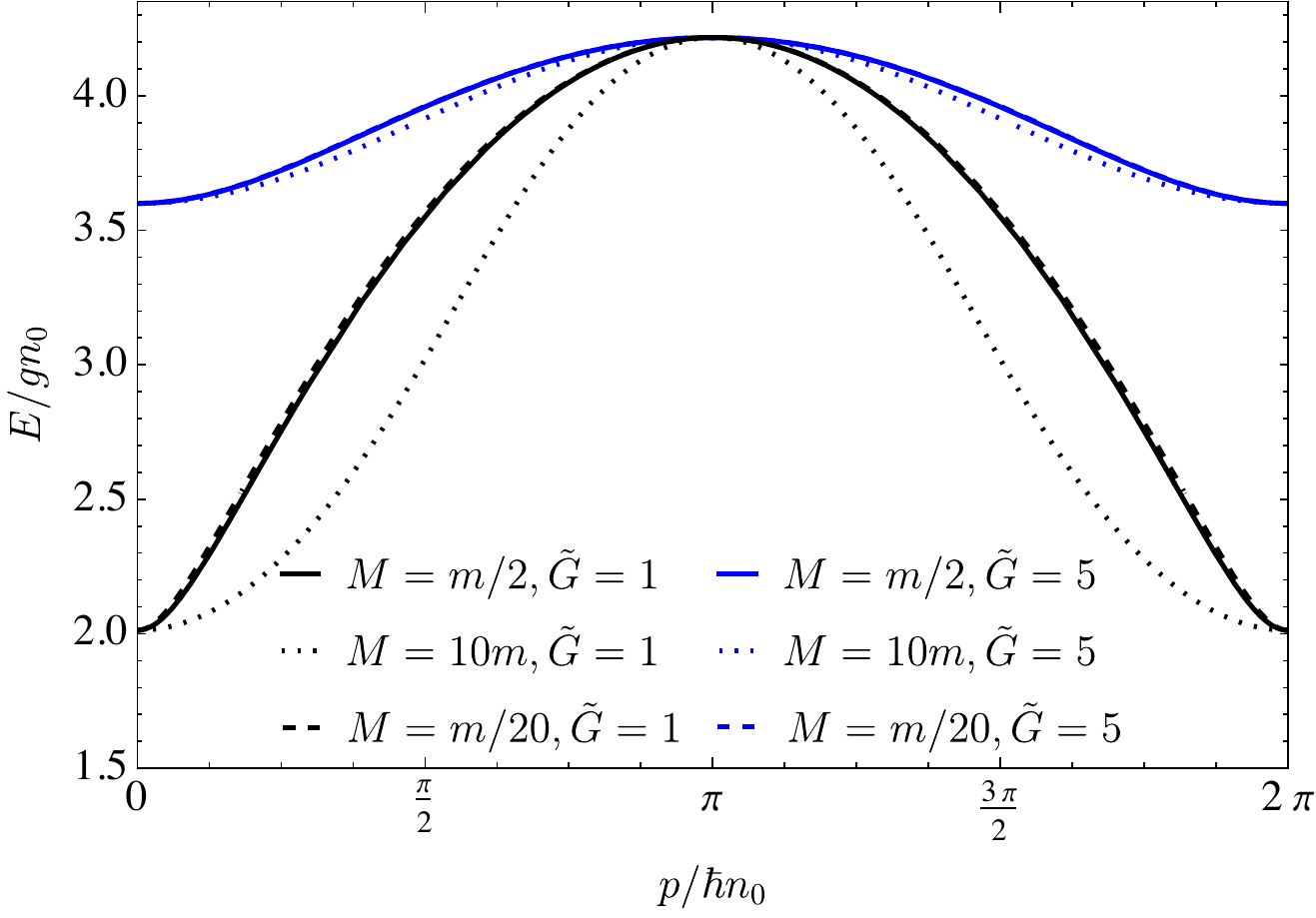}
    \caption{Energy of the polaron for the solution of \eq{eq:mean-field2} as a function of the initial impurity momentum for $M=m/2$ (solid line), $M=10m$ (dotted line), and $M=m/10$ (dashed line) for two different values of the dimensionless impurity coupling $\tilde{G}$. Here $\gamma = 0.1$.}
    \label{fig:energy2}
\end{figure}

Next we consider the stationary solution of \eq{eq:mean-field2}. Since \eq{eq:mean-field2} differs from \eq{eq:mean-field1} only in the term containing the second spatial derivative, all equations from Sec.~\ref{sec:stationary1} apply to \eq{eq:mean-field2} once $1+m/M$ is replaced by $1$. Then, the critical velocity defined by modified \eq{eq:vc} becomes $M$-independent.  This is illustrated in Fig.~\ref{fig:vf2}, which is in stark contrast to Fig.~\ref{fig:vf}. Besides, the maximal polaron energy does not depend on $M$ and $\tilde{G}$, as portrayed in Fig.~\ref{fig:energy2}.
The difference between the cases $M=m/20$ and $M=m/2$ is almost not visible at Fig.~\ref{fig:energy2}, especially at stronger impurity coupling. However, the  physics of a heavy impurity, discussed in the previous subsection, remains unchanged. 

The binding energy of the impurity, $E_p(0)$,  in the limit of strong coupling $\tilde{G}=G/\hbar v\gg 1$ is $E_p(0)=4g n_0/3\sqrt{\gamma}$ for the solution of \eq{eq:mean-field2}. The same quantity for the solution of \eq{eq:mean-field1} 
is given by \eq{eq:PolaronEnergy} evaluated at $p=0$ and it is bigger by a factor of $\sqrt{1+m/M}$. However, 
note that Refs.~\cite{VolosnievPolaron,Pastukhov2019, PhysRevResearch.2.033142}  report successful comparison of the binding energy for the solution of \eq{eq:mean-field1} with known numerical results \cite{PhysRevA.95.023619,Grusdt_2017} as a function of the impurity strength, including the case of strongly coupled impurity at $M\sim m$ where the difference between two expressions is considerable.

\begin{figure}
    \centering
    \includegraphics[width= 0.95\columnwidth]{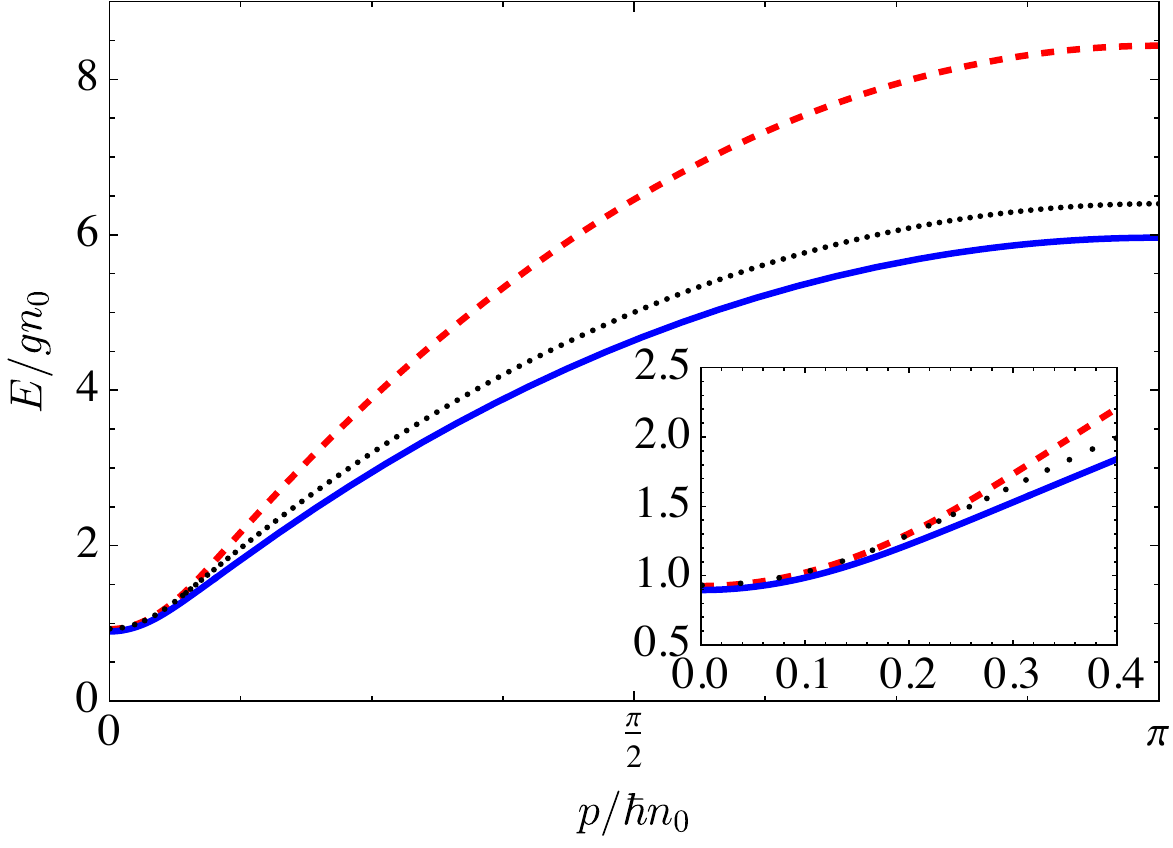}
    \caption{Polaron energy as a function of the momentum for the Yang-Gaudin model, $M=m$ and $g=G$. Here $\gamma=1/20$. The dotted line shows the exact result, the dashed red line is
 \eq{eq:PolaronEnergy}, while the blue solid line is the energy dependence on $p$ for the solution of \eq{eq:mean-field2}. The inset shows the energies for small momenta.}
    \label{fig:comparison}
\end{figure}

The impurities can be fabricated by preparing some atoms in a different hyperfine state with
respect to the majority of atoms \cite{palzer2009quantum}. Thus $M=m$ is an experimentally relevant case. Furthermore, for $G=g$ and $M=m$, the Hamiltonian (\ref{eq:H}) describes the Yang-Gaudin model. The latter is integrable, and admits exact results for the energy \cite{ZvonarevPhysRevB.80.201102,fuchs2005spin, Zoran_YangG}. 
After expanding our results for the polaron energy in small $G/\hbar v=\sqrt{\gamma}$ for the solution of \eq{eq:mean-field2}, we obtain the leading order term of the energy dependence on momentum of the exact result, while  \eq{eq:PolaronEnergy} gives a result bigger by a factor $\sqrt{2}$ for the momentum above $p^* \sim \hbar n_0 \sqrt{\gamma}$ where the second stationary solution is the ground state.
The origin of this multiplicative factor is $\sqrt{1+m/M}=\sqrt{2}$. In Fig.~\ref{fig:comparison} we show the comparison of energies for the solutions of \eq{eq:mean-field1}, \eq{eq:mean-field2} and the exact result. For the parameters used for Fig.~\ref{fig:comparison}, the crossover from the first to the second solution happens at $p^*\approx 0.35\hbar n_0$ for \eq{eq:mean-field1} and $p^*\approx 0.32 \hbar n_0$ for \eq{eq:mean-field2}. The analysis presented in this paragraph suggests that the mean-field \eq{eq:mean-field2} describes the physics of this model better than \eq{eq:mean-field1} for a nonzero system momentum $p\geq p^*$.

\section{Time-evolution after a quench\label{sec:quench}}

In this section we consider the following problem. At initial time the impurity is free and has a finite momentum, while the bosons are in their ground state. Then, the impurity-boson coupling is switched on, or the impurity is injected into the system of bosons, and we follow the relaxation starting from this far-from-equilibrium initial state. 
The system is assumed to be at zero temperature. 

We solve  \eq{eq:mean-field1}  with periodic boundary conditions numerically. The discretization of time and space is done using the protocol of conservative finite difference scheme implemented fully implicitly, and the obtained system of nonlinear algebraic equations is solved iteratively \cite{GPE_discretization}. 
In the discretization of the drift term, a first-order upwind scheme has been implemented to respect the causality of the propagation of the impurity \cite{upwind_scheme}. 
The system size is taken sufficiently long, such that the density waves do not reach the boundaries of the system during the considered time-evolution. 

\begin{figure}
	\includegraphics[width=0.49\columnwidth]{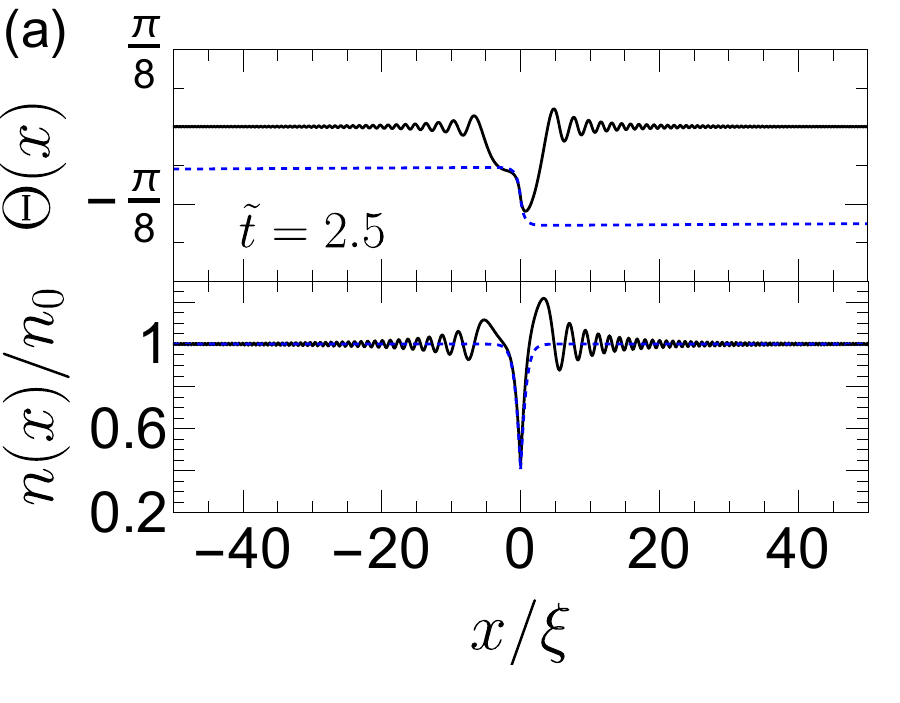} 
	\includegraphics[width=0.49\columnwidth]{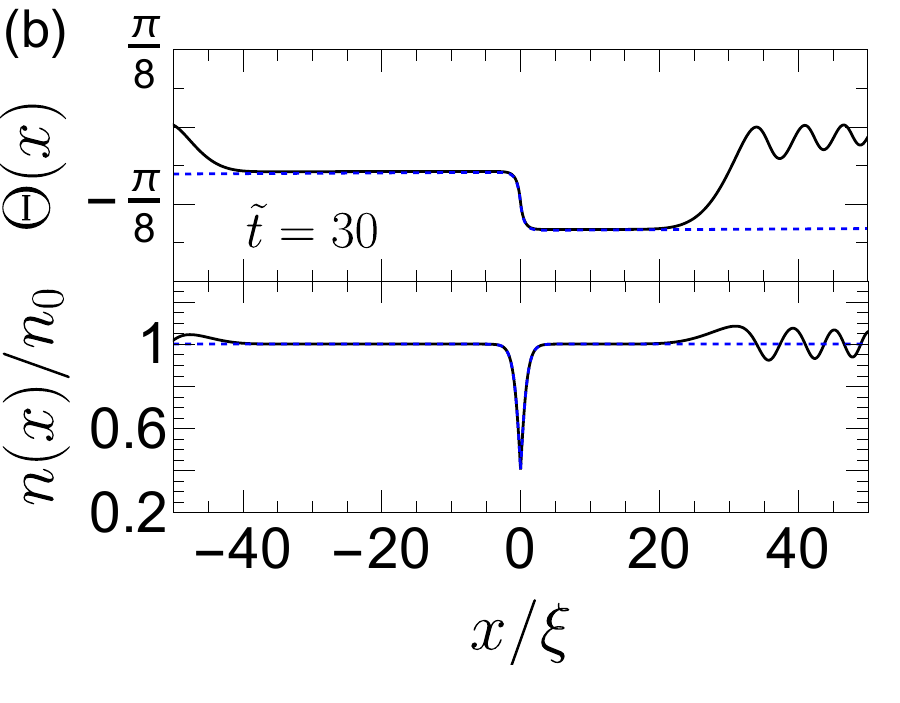}
	\caption{Phase and boson density profile in the frame co-moving with the impurity for $V_0=0.5v$, $M=3m$, $\tilde{G}=1$ and $\gamma = 0.1$ at (a) $\tilde{t}=2.5$ and (b) $\tilde{t}=30$. The dashed line shows the density and the phase profile of the stationary state \eq{eq:MFsolution} for numerically obtained final impurity velocity.}\label{fig1}
\end{figure}

\begin{figure*}
	\includegraphics[width=0.435\linewidth]{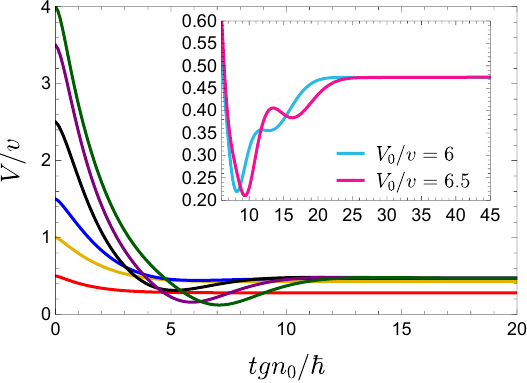} \phantom{aaaaaaa}
	\includegraphics[width=0.45\linewidth]{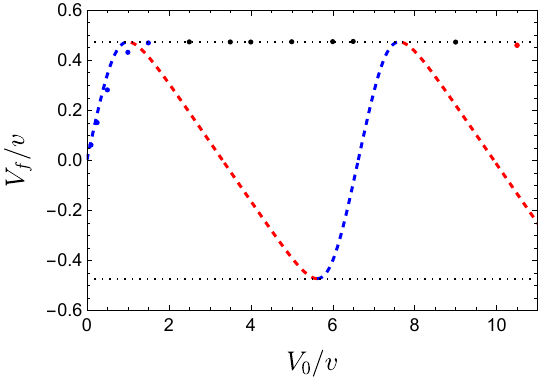} 
	\caption{(\emph{left}) Time-evolution of the impurity velocity for $M=3m$ and $\tilde{G}=1$ for different initial velocities. Here $\gamma = 0.1$. The inset shows the time evolution of the impurity velocity for the same parameters, but for higher initial velocities $V_0=6v$ and $V_0=6.5v$.
(\textit{right}) For the same parameters, the final dimensionless impurity velocity $V_f/v$ as a function of the initial dimensionless impurity velocity $V_0/v$. The dashed lines and the dots correspond to the analytic prediction (\ref{eq:momentum}) and the numerically obtained values, respectively. The blue and the red colours correspond to the 1st and 2nd stationary solutions, respectively. We used the black dots for the impurity velocity being equal to the critical velocity, where the two stationary states coincide.}
\label{fig2}
\end{figure*}

\begin{figure*}
	\includegraphics[width=0.97\columnwidth]{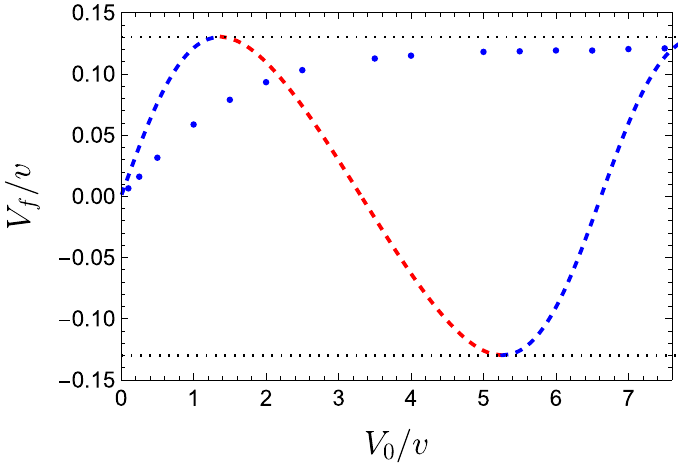} \phantom{aaaaaaa}
	\includegraphics[width=0.95\columnwidth]{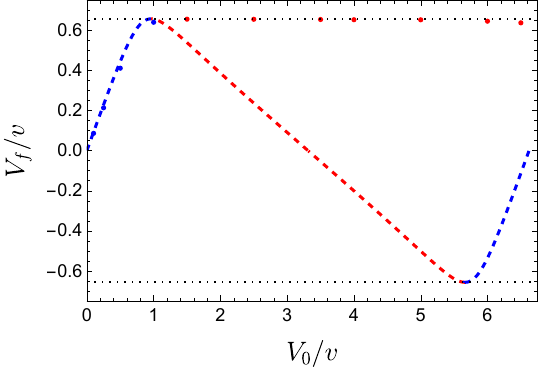} 
	\caption{Final impurity velocity as a function of the initial impurity velocity  for $M=3m$ and $\gamma=0.1$ for  (\emph{left}) $\tilde{G}=5$ and (\emph{right}) $\tilde{G}=0.5$. The dashed lines and the circles correspond to the analytic prediction (\ref{eq:momentum}) and the numerically obtained values, respectively. The blue and the the red colours correspond to the 1st and 2nd stationary solutions, respectively. }
\label{dynam}
\end{figure*}

\begin{figure}
	\includegraphics[width=0.95\linewidth]{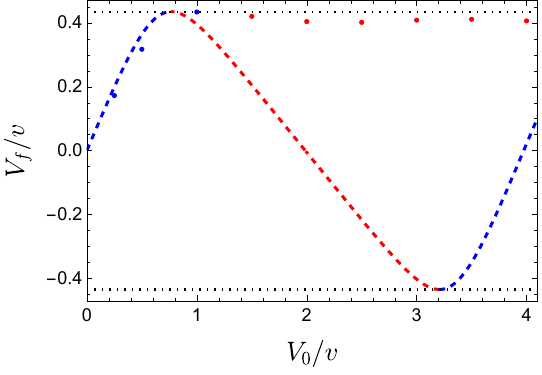} 
	\caption{Final impurity velocity $V_f/v$ as a function of the initial impurity velocity $V_0/v$ for $M=5m$, $\tilde{G}=1$ and $\gamma=0.1$. The dashed lines and the circles correspond to the analytic prediction (\ref{eq:momentum}) and the numerically obtained values, respectively. The blue and the red colours correspond to the 1st and 2nd stationary solutions, respectively.}
\label{G1M5}
\end{figure}

\subsection{Cherenkov radiation} 

The following general physical picture emerges in the process of relaxation.  The impurity perturbs the boson density profile at its position and triggers the emission of density dispersive shock waves \cite{PhysRevA.69.063605,PhysRevA.74.023623,atoms10010003} shown in Fig.~\ref{fig1}. The fronts of the density waves move away from the impurity with a velocity $v\sqrt{1+m/M}$ in the laboratory frame, carrying away some momentum as well as background particles and leaving the hole, see Fig.~\ref{fig1}. 
The hole evolves in time into the stationary state (\ref{eq:MFsolution}). However, as some momentum is imparted into the density waves, the analytic prediction (\ref{eq:momentum}) for the impurity final velocity fails in a general case, as will be explained in the next section. 
In Fig.~\ref{fig1}, we show the boson density relative to the impurity position, $|\Psi_0(x,t)|^2$, and in the following we refer to it as the boson density. We also show the time-evolution of the phase of the bosons. For the parameters shown in Fig.~\ref{fig1}, the waves are faster than the impurity, see Fig.~\ref{fig2}, and therefore there is a density disturbance on each sides of the impurity. The density profile is asymmetric. Bigger the initial impurity velocity is, bigger the disturbance amplitude on the right hand side of the impurity is.  In the process of Cherenkov radiation, the impurity slows down, but does not come to a full stop as shown in Fig.~\ref{fig2}. This is a hallmark of superfluidity. The background bosons exert no friction force on a slow impurity with velocity $V\leq v_c$ in equilibrium conditions at zero temperature \cite{landau+49,castro_neto1996dynamics}. Thus the final velocity is smaller or equal to the critical one.
Note that at nonzero temperatures, a slow impurity experiences a finite friction \cite{MyPRLdissipative}.

\subsection{Dynamical crossover and its dependence on $M/m$ and $\tilde{G}$ \label{sec:crossover}}

The analytic solution (\ref{eq:MFsolution}), which we refer to as the stationary one, is obtained without any knowledge about the initial conditions and the quench of the impurity-boson coupling constant. Thus many questions remain unanswered. What is the final impurity velocity for a given initial one? How does it depend on the system parameters, such as $M/m$, $\tilde{G}$ and $V_0/v$? We will address these issues in this section.

First, we study the dynamics by fixing the impurity mass and the coupling constant and varying its initial velocity. We start our analysis for not very heavy impurities. The relaxation process is slower for initially faster impurities, as shown in Fig.~\ref{fig2}. Furthermore, for sufficiently big initial momentum, the impurity velocity exhibits a characteristic minimum in time before increasing and reaching a constant stationary value \cite{knap2014quantum,Will_2023}. For even higher initial momenta,  the velocity displays more complex non-monotonic behavior in time with appearance of additional local minima, as shown in the inset of Fig.~\ref{fig2}, with the tendency to develop the oscillations, see Sec.~\ref{sec:flutter}.

The most striking effect is that the final impurity velocity takes a constant value $V_s$ at sufficiently big initial momenta and strongly deviates from the analytic prediction (\ref{eq:momentum}), see Fig.~\ref{fig2}. The depletion of the boson density in the vicinity of the impurity follows from  \eq{eq:MFsolution}, but the final impurity velocity is not given by Eq.~(\ref{eq:momentum}) for a given $V_0$.
The reason of this discrepancy lies in the fact that the analytic solution (\ref{eq:MFsolution}) does not take into account the emitted density waves \cite{atoms10010003}. The momentum (\ref{eq:momentum}) includes three terms: the momentum of the local hole, of the impurity, and the contribution caused by the phase gradient that takes place over the whole system. We stress that all three contributions are perturbed by the density waves. In Fig.~\ref{fig1}, we show how the phase profile of the bosons is affected by the density waves.

The underlying mechanism for the final impurity velocity saturation is the following. Increasing the initial impurity velocity, the momentum carried by the dispersive shock waves increases, while the final impurity velocity remains constant and locally the system remains in the very same stationary state for a large interval of initial velocities. In Fig.~\ref{fig2}, the impurity final velocity is constant within the numerical error for $10.5>V_0/v\geq2.5$.  At $V_0=10.5v$, we observe a drop of approximately $2\%$ of $V_s$. 

\begin{figure}
	\includegraphics[width=0.95\linewidth]{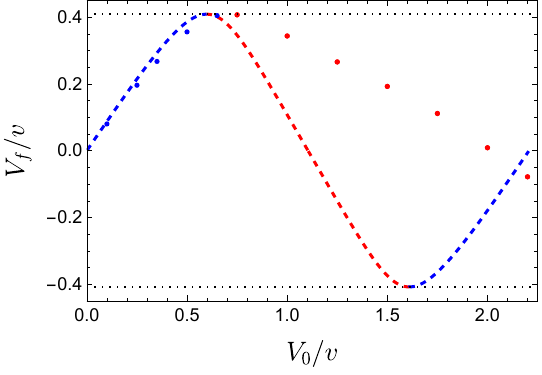} 
	\caption{Final impurity velocity $V_f/v$ as a function of the initial impurity velocity $V_0/v$ for $M=9m$, $\tilde{G}=1$ and $\gamma=0.1$. The dashed lines and the circles correspond to the analytic prediction (\ref{eq:momentum}) and the numerically obtained values, respectively. The blue and the red colours correspond to the 1st and 2nd stationary solutions, respectively.}
\label{M9m}
\end{figure}

On the other hand, at very low momenta $p=M V_0$, the deviation of the final impurity velocity with respect to the one obtained from Eq.~(\ref{eq:momentum}) is less important. Increasing $p$, the deviation increases.
Moreover,  for bigger coupling constants $\tilde{G}$, the mismatch is important even at low $p$, as illustrated in Fig.~\ref{dynam}. The crossover into a constant impurity velocity depends strongly on $\tilde{G}$, and it is wide in the strong coupling regime. The saturation value of the velocity, $V_s$, depends on $\tilde{G}$ and $M/m$. It can be written as $V_s=v \tilde{V}_s(\tilde{G},M/m)$, where $\tilde{V}_s$ is a function of two dimensionless parameters. We point out that in a general case $V_s$ differs from the critical velocity (\ref{eq:vc}). 

For smaller impurity coupling constant $\tilde{G}=0.5$ shown in Fig.~\ref{dynam}, the velocity drops for approximately $1\%$ and $2\%$ of $V_s$ at $V_0=6$ and $V_0=6.5$, respectively. For $1.5\leq V_0/v< 6$, $V_f$ is constant within the numerical error. This might signal that the region of constant impurity velocity shrinks down with decreasing $\tilde{G}$. Nevertheless, we point out that in a very large and experimentally relevant  interval of initial velocities, the system remains locally in the very same stationary state with a constant impurity velocity. 

For heavier impurities, $V_f$ also deviates strongly from the analytic prediction (\ref{eq:momentum}), see Fig.~\ref{G1M5}. The system crosses over from the first to the second stationary state, and then the final velocity remains almost constant. It actually displays small variations over the wide interval of initial velocity, for $4\geq V_0/v\geq 2$ they are approximately $1.7\%$ of the minimal $V_f$ obtained in that interval.

For a heavy impurity, the impurity relaxation dynamics is very different than for a lighter one with the same $\tilde{G}$ and the same initial momentum. In Fig.~\ref{M9m}, we show the final impurity velocity as a function of the initial one for $M=9m$.  Contrary to a light impurity, a heavy impurity with the very same initial momentum succeeds to further locally deplete the boson density. As a result the impurity velocity decreases as a function of $p$ while being in the second stationary state.  Thus, a heavy impurity can change its direction of motion, as shown in Fig.~\ref{M9m}. 

\begin{figure*}[ht]
\centering
	\begin{tabular}{cc}
 \includegraphics[width=0.9\columnwidth]{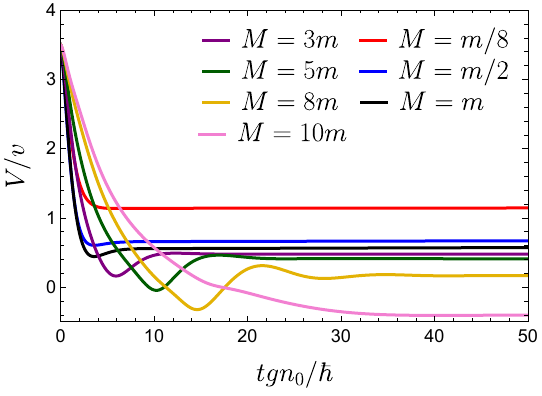} \phantom{aaaaaa}   
 \includegraphics[width=0.9\columnwidth]{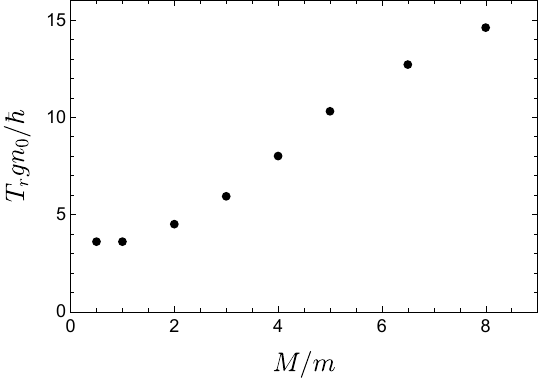} \\
\end{tabular}
\caption{(\emph{left}) Time-evolution of the impurity velocity for the initial velocity $V_0=3.5 v$ and $\tilde{G}=1$ for different values of impurity masses. Here  $\gamma = 0.1$. (\emph{right}) Characteristic time scale $T_r$ as a function of $M/m$ for the same parameters. 
		 } \label{figV2.5}
\end{figure*}

\subsection{Mass dependence of the characteristic relaxation time \label{sec:Mdependence}}

Next, we study the dynamics of impurities having different masses, but the same initial velocity and the coupling constant, see Fig.~\ref{figV2.5}.
For impurities heavier than the background bosons, the relaxation slows down considerably. The time $T_r$ where the impurity velocity reaches its minimal value is one characteristic relaxation time-scale. The latter is shown as a function of $M/m$ in Fig.~\ref{figV2.5}. In a large interval of masses with $M\gtrsim2m$, $T_r$ is proportional to $M/m$. 

All obtained final impurity velocities in Fig.~\ref{figV2.5} are below the critical velocity (\ref{eq:vc}), that itself satisfies $v_c\leq v\sqrt{1+m/M}$. Note that the sound velocity that follows from \eq{eq:mean-field1} is $v\sqrt{1+m/M}$. 
As a result, the final velocity for a sufficiently light impurity can exceed $v$, as shown in Fig.~\ref{figV2.5}. As discussed in Sec.~\ref{sec:mean-field}, this problem does not exist for \eq{eq:mean-field2} that gives the correct sound velocity. Furthermore, note that the quantum fluctuations are more relevant for a light impurity.

The impurity time-evolution strongly depends on its mass, showing very different regimes. In Fig.~\ref{figV2.5}, the stationary state is the first state for $M<5m$. Then at $M=5m$ and $M=8m$ the second stationary state occurs, and the dynamics at intermediate time scales becomes different. The impurity velocity exhibits damped oscillations. We study this phenomenon in Sec.~\ref{sec:flutter}. For heavier impurities, at $M=10m$, we observe a soliton emission and absence of the velocity oscillations. The stationary state of the system is given by the first state. Furthermore, the impurity changes the direction of motion \cite{Will_2023}. Although its initial velocity is positive, at $M=10m$, the final velocity is negative. Note that in this case, the impurity mass is smaller than the critical one. For very heavy impurities with $M>M_c$, a very different dynamics arises, as shown in Sec.~\ref{Sec:M>Mc}.

\subsection{Temporal impurity velocity oscillations \label{sec:flutter}}

\begin{figure*}
	\centering
	\includegraphics[width=0.9\columnwidth]{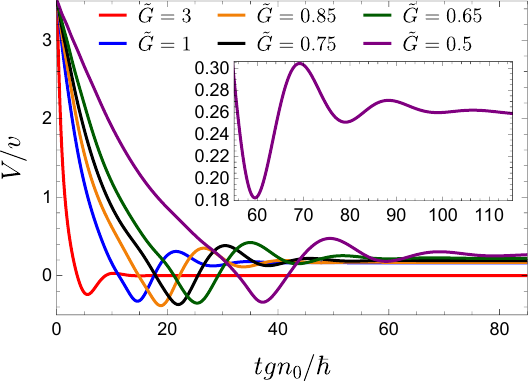} \phantom{aaaaaaa}
	\includegraphics[width=\columnwidth]{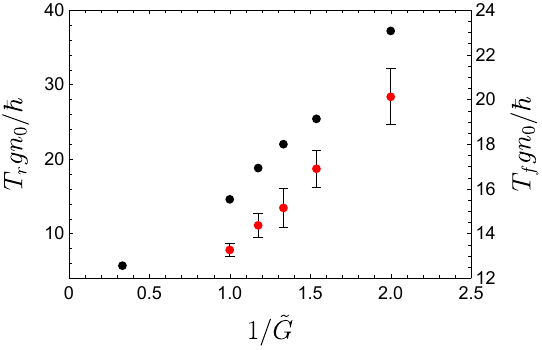} 
	\caption{(\emph{left}) Impurity velocity evolution for $M=8m$, $V_0=3.5v $, and $\gamma=0.1$ for different impurity coupling strengths.  The inset shows the long-time behaviour of the impurity velocity for $\tilde{G}=0.5$. (\emph{right})  The red (black) points correspond to the time period  $T_f$ (relaxation time $T_r$ where the velocity reaches its minimal value) as a function of $1/\tilde{G}$. }
	\label{fig:Gdependence}
\end{figure*}

\begin{figure*}
	\centering
	\includegraphics[width=2.1\columnwidth]{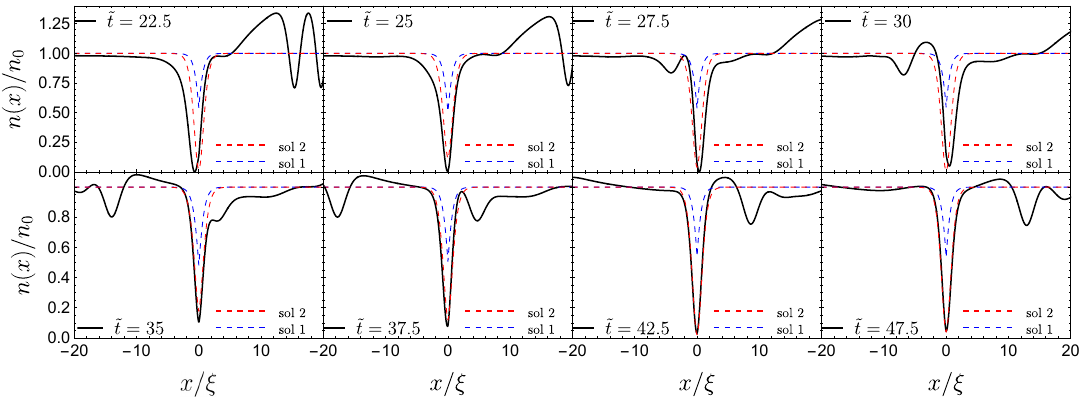} 
	\caption{Boson density profile as a function of time for $M=8m$, $V_0=3.5v $, $\gamma=0.1$ and $\tilde{G}=0.65$. The curve $V(t)$ is shown in Fig.~\ref{fig:Gdependence}. The blue and red lines correspond to the density profile of the 1st and 2nd stationary solutions for the impurity velocity at given time, respectively.}
	\label{oscillations}
\end{figure*}

In Fig.~\ref{fig:Gdependence}, we show a strong influence of the impurity coupling strength $\tilde{G}$ on the dynamics. Weaker the coupling to the bath is, slower the relaxation is. The characteristic relaxation time-scale is shown as a function of $\tilde{G}^{-1}$ in Fig.~\ref{fig:Gdependence}.  

Even more important is the appearance of a new dynamical regime for the case of an impurity having a sufficiently big initial momentum, shown in Fig.~\ref{fig:Gdependence}. After a rapid drop, the impurity velocity exhibits damped oscillations in time before converging to its final stationary value. We stress that the oscillations become more pronounced and longer-lived as $\tilde{G}$ decreases. On the contrary, for a sufficiently big impurity coupling constant the oscillations disappear.  Besides, Fig.~\ref{fig:Gdependence} suggests  that despite nonmonotonic dependence on $\tilde{G}$ of  the minimal impurity velocity reached in time, the final velocity $V_f$ is a decreasing function of $\tilde{G}$. 

Next we consider the underlying physical mechanism for the above mentioned observations. The force exerted by the bosons on the impurity is given by the gradient of the potential, and reads as
\begin{align}\label{eq:friction}
F(t)=&\int \mathrm{d} x \Psi_0^*(x,t) \partial_x\left[G \delta(x)\right] \Psi_0(x,t)\notag\\=&-\frac{G}{2}\left[\partial_x n(x,t)|_{x=0^-}+\partial_x n(x,t)|_{x=0^+} \right]
\end{align}
in the reference frame co-moving with the impurity. In the following discussion, we will denote $\left[\partial_x n(x,t)|_{x=0^-}+\partial_x n(x,t)|_{x=0^+} \right]/2=\partial_x n(0,t) $.
In Fig.~\ref{oscillations}, we show the time evolution of the boson density in the vicinity of the impurity for the data shown in Fig.~\ref{fig:Gdependence} for $\tilde{G}=0.65$. We focus on the time evolution once the local hole in the density has been formed and the impurity velocity oscillations appear. 
At $\tilde{t}=22.5$, the impurity has already changed the direction of motion and its velocity is negative. The density profile in Fig.~\ref{oscillations} shows that the hole is in front of the impurity and that the density of bosons satisfies $\partial_x n(0)>0$.  Thus, the force (\ref{eq:friction}) is actually negative, $F<0$, and leads to the acceleration of the impurity. This is consistent with the velocity curve shown in Fig.~\ref{fig:Gdependence}, where we see that the impurity accelerates at $\tilde{t}=22.5$. At $\tilde{t}=25$, the absolute value of the impurity velocity approaches its local maximal value, while $\partial_x n(0)$ and the friction force approach zero. Afterwards, the oscillating depletion cloud comes on the other side of the impurity, the force changes the sign and leads to the impurity deceleration. This is seen in the density profile given at $\tilde{t}=27.5$. Note that at that moment the impurity velocity  $V$ is negative. As a result, the impurity slows down, comes to a full stop, changes the direction of motion and then accelerates. This is seen at $\tilde{t}=30$, where $V>0$ and $F>0$. At $\tilde{t}=35$, the impurity reaches its local maximal velocity and the friction becomes zero. 
This process repeats itself in time as long as the maximal depletion of the boson density is off the impurity position.  The underlying cause is the effective attraction (\ref{eq:friction}) between the hole and repulsive impurity, as well as the inertial effects. However, each collision of the hole and the impurity triggers the emission of density waves, as seen at $\tilde{t}=27.5$ on the left hand side and $\tilde{t}=35$ at the right hand side of the impurity. As a result, the dressed impurity loses its energy, the oscillations become damped, and system converges into a stationary state (\ref{eq:MFsolution}) where the force (\ref{eq:friction}) vanishes.
Since the total momentum is conserved and the impurity momentum oscillates, the momentum  of the bosons oscillates as well. This is reflected in the change of the shape and the depth of the hole, that are visible in Fig.~\ref{oscillations}.

We stress that even though for each final impurity velocity there are two possible stationary states, during the velocity oscillations the system does not oscillate between them, but stays in the vicinity of one state, as shown in Fig.~\ref{oscillations}. In Fig.~\ref{fig:Gdependence}, it is actually the second stationary state that is reached for all shown values of $\tilde{G}$. 

Decreasing the impurity coupling constant, the attraction between the hole and the impurity diminishes, resulting in bigger amplitude and bigger period of the oscillations. The latter is shown in Fig.~\ref{fig:Gdependence} as a function of $\tilde{G}$. The period and its error are calculated by taking the distance in time between the neighbouring minima (maxima) and evaluating their mean and the standard deviation.
Furthermore, we stress that the oscillations become more pronounced by increasing the strength of the repulsion between bosons, while keeping the other parameters fixed. 

The phenomenon of slowly decaying velocity oscillations of fast impurities injected into a system of strongly interacting one-dimensional bosons was dubbed quantum flutter  \cite{QFlutterNature,knap2014quantum}. 
The underlying mechanism described above of the back and forth motion of the hole around the impurity takes place also in the Tonks-Girardeau model with an impurity \cite{QFlutterNature}.
However, in Ref.~\cite{QFlutterNature}, it has been
conjectured that the phenomenon cannot be described by a hydrodynamic approach and the Gross-Pitaevskii equation. The authors interpreted it as coherent oscillations between the Lieb's type-II excitation and the polaron state, both taken at momentum $p=\pi \hbar n_0$. 
At weak repulsion between the bosons, the Lieb's type-II excitation at $p=\pi \hbar n_0$ is a dark soliton \cite{ishikawa1980}, while the polaron configuration at $p=\pi \hbar n_0$ consists of the impurity with zero velocity, sitting at the center of a dark soliton. Fig.~\ref{oscillations} shows that only in the initial stage of the oscillations the local boson density depletion is complete, as the impurity velocity approaches its first local maximal value and later on, it differs considerably from a dark soliton.

\begin{figure}
	\centering
	\includegraphics[width=\columnwidth]{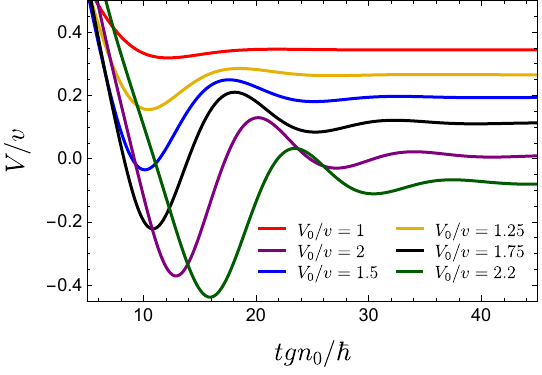} 
	\caption{Impurity velocity time evolution for $M=9m$, $\tilde{G}=1$ and $\gamma=0.1$ for different initial velocities. }
	\label{Voft9m}
\end{figure}

Fig.~\ref{Voft9m} displays the impurity velocity as a function of time for different $V_0$ at fixed mass $M=9m$. We observe the emergence of the impurity velocity oscillations as $V_0$ is increased.  Moreover, we note a weak dependence on $V_0$ of the characteristic relaxation time where the velocity as a function of time reaches its minimal value, as was the case also for a light impurity shown in Fig.~\ref{fig2}. On the other hand, the time for a complete relaxation into a stationary state now increases considerably with $V_0$.
Note that for a very big initial momentum, a soliton emission takes place leading to the absence of the oscillations, as shown in fig.~\ref{figV2.5} for $M=10m$. 

\begin{figure}
	\centering
	\includegraphics[width=\columnwidth]{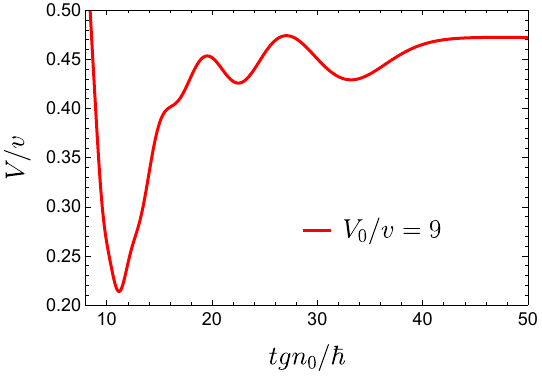} 
	\caption{Impurity velocity as a function of time for $M=3m$, $\gamma=0.1$, $\tilde{G}=1$ and $V_0=9v$. }
	\label{flutter:mass}
\end{figure}

We have shown that the ratio $M/m$ has a very important influence on the impurity dynamics after a quench, see Secs.~\ref{sec:crossover} and \ref{sec:Mdependence}. Regarding the temporal oscillations considered in this section, we point out that decreasing the impurity mass while keeping all the parameters the same, including the initial impurity momentum, the impurity velocity displays different kind of oscillations shown in Fig.~\ref{flutter:mass}. In contrast to Figs.~\ref{Voft9m} and \ref{fig:Gdependence}, the oscillations are not damped, but their amplitude increases in time, before a stationary state is reached. These oscillations are also shown in the inset of Fig.~\ref{fig2}. Moreover, the underlining mechanism is different. The aforementioned accompanying oscillations of the hole around the impurity are absent here, and  the maximal depletion of the boson density is situated at the impurity position during the impurity velocity oscillations. Thus, the relaxation of a heavy fast impurity differs from a light impurity with the same momentum.

\subsection{Very heavy impurity, $M>M_c$\label{Sec:M>Mc}} 

\begin{figure*}
   \centering
    \includegraphics[width=0.43\linewidth]{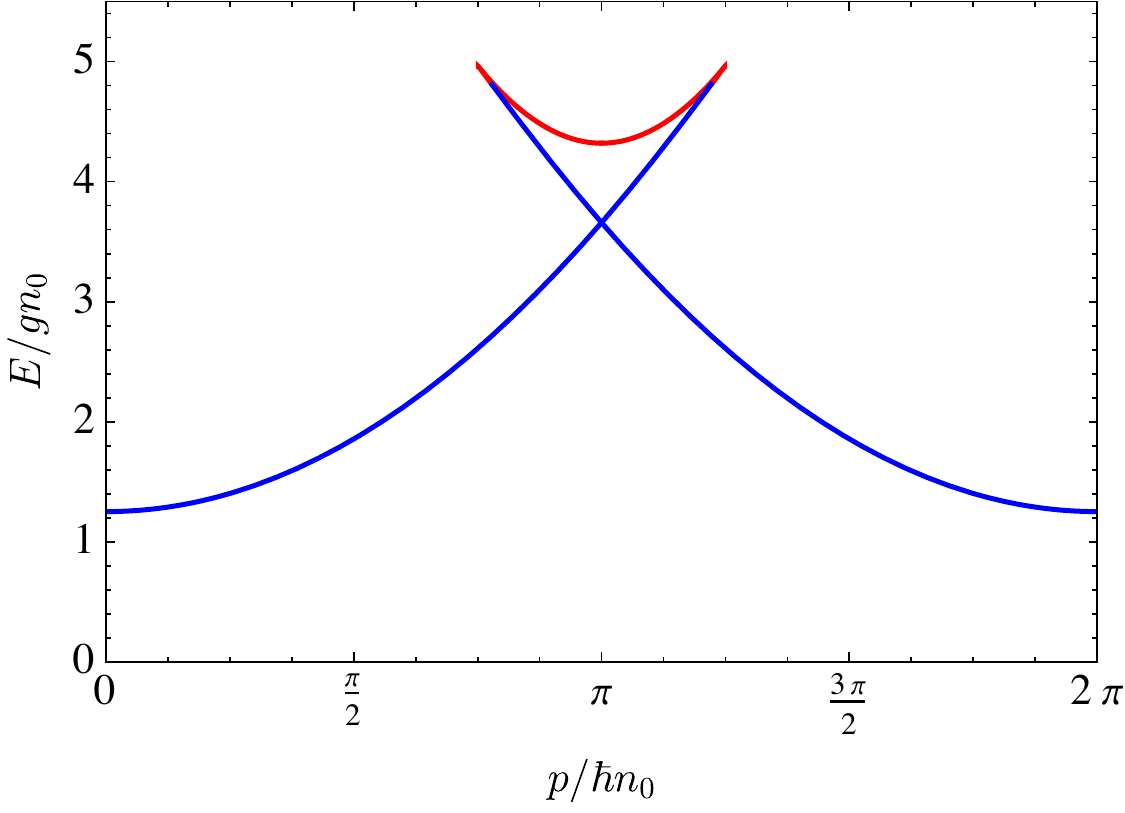}\phantom{aaaaa}
     \includegraphics[width=0.45\linewidth]{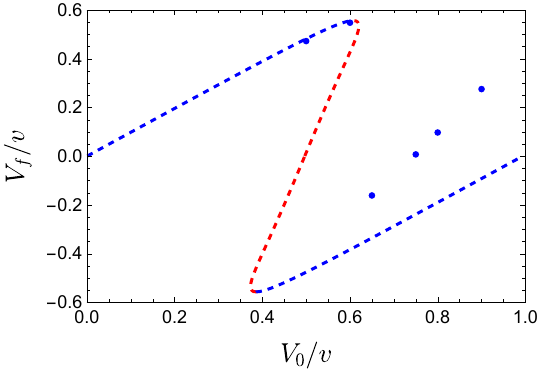}
    \caption{(\textit{left})  Polaron energy dispersion (\ref{eq:PolaronEnergy}) for $\tilde{G}=0.5,M=20 m$ and $\gamma=0.1$. (\textit{right}) Final impurity velocity as a function of the initial impurity velocity for the aforementioned parameters. The dashed lines and the circular points denote the analytic and the numerical results, respectively. The blue and the red colours correspond to the 1st and 2nd stationary states, respectively.}
    \label{fig:aboveMc}
\end{figure*}
\begin{figure*}
    \centering
    \includegraphics[width=\columnwidth]{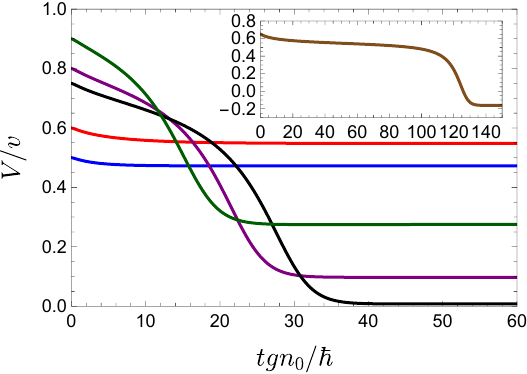}\phantom{aaa}
    \includegraphics[width=\columnwidth]{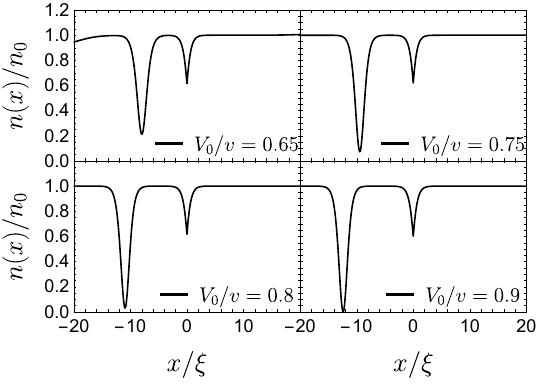} 

    \caption{(\textit{left}) Impurity velocity as a function of time for different initial velocities.  The inset shows the case of $V_0=0.65 v$. (\textit{right}) Boson density profile reveals the presence of a soliton. The parameters are the same as in Fig.~\ref{fig:aboveMc}: $\tilde{G}=0.5, M=20m$ and $\gamma=0.1$.}
    \label{fig:Single_soliton}
\end{figure*}

Here we investigate the time-evolution of an impurity with a mass exceeding the critical mass and some manifestations of its energy dispersion. The latter is discussed in Sec.~\ref{sec:stationary1}.   In Fig.~\ref{fig:aboveMc}, we show numerical results for the obtained stationary state and the final impurity velocity as a function of the initial velocity. As expected, the second stationary state does not occur.
Increasing the impurity initial velocity just beyond the termination point of the metastable blue branch, the system relaxes into the true ground state. We point out that the relaxation is considerably slowed down, see the case of $V_0=0.65v$ in the inset of Fig.~\ref{fig:Single_soliton} where we provide the impurity velocity as a function of time. Moreover, the impurity changes its direction of motion. Note that in this process, apart from the density waves, also a soliton emission takes place. 

The time-evolution of the impurity velocity for different initial velocities, and the corresponding boson density profiles are shown in Fig.~\ref{fig:Single_soliton}. Increasing the initial velocity from $V_0=0.65v$ to $V_0=0.9v$, the energy difference between the initial state and of the locally obtained polaron state, given by Eq.~(\ref{eq:PolaronEnergy}), increases. As a result, the emitted solitons become more energetic and characterised by a bigger boson density depletion, as shown in Fig.~\ref{fig:Single_soliton}.

In Fig.~\ref{fig:aboveMc},  we observe that the system remains stuck in the metastable state during the time we have followed its dynamics. However, once the quantum fluctuations are taken into account, the system can tunnel from the metastable to the true ground state, and the emission of solitons can take place for initial velocities smaller than the initial velocity of the termination point of the blue branch. The latter is not the critical velocity for $M>M_c$ once the quantum fluctuations are taken into account, as explained in Sec.~\ref{sec:stationary1}.

\subsection{Relaxation dynamics given by \eq{eq:mean-field2} \label{sec:dynamics2}}

\begin{figure*}[ht]
    \centering
    \includegraphics[width=\columnwidth]{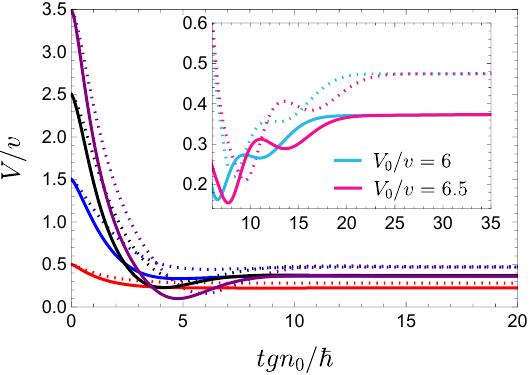}\phantom{aaa}
    \includegraphics[width=\columnwidth]{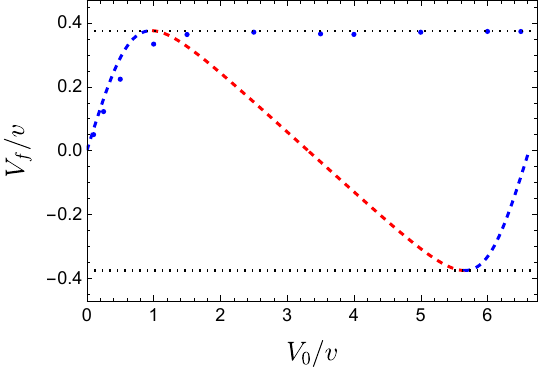}
    \caption{ (\emph{left}) Impurity velocity as a function of time for  \eq{eq:mean-field2} and \eq{eq:mean-field1} is shown in full and dotted line, respectively. (\emph{right}) Final impurity velocity as a function of the initial one for  \eq{eq:mean-field2}. For both figures  $\gamma=0.1, \tilde{G}=1$, and $M=3m$.}
    \label{dynamNEWGP}
\end{figure*}

Next, we study the dynamics described by \eq{eq:mean-field2}. The conclusions regarding the relaxation of an impurity with the mass $M>M_c$, as well as for the impurity velocity oscillations remain the same since they concern heavy impurities, where the difference between the two equations of motion is negligible. 
We thus focus on a possible dynamical crossover and the characteristic time scales of relaxation.

In Fig.~\ref{dynamNEWGP}, we consider the case of $M=3m$ and $\tilde{G}=1$ for different initial velocities. 
We see that the dynamical crossover into almost constant final impurity velocity takes place. For the initial velocities  $V_0\geq 2.5v$ considered in Fig.~\ref{dynamNEWGP}, the variations of the final velocity are approximetely  $2.5\%$ of the minimal final velocity obtained. They are bigger than for the dynamics described by \eq{eq:mean-field1} shown in Fig.~\ref{fig2} for the same parameters.

Note that the impurity velocity as a function of time, obtained by solving \eq{eq:mean-field1} for the very same parameters, is shown in Fig.~\ref{dynamNEWGP} by the dotted line. We thus see that the initial stage of the relaxation is faster, the drop of the velocity is bigger and as a result $V_f$ obtained by \eq{eq:mean-field2} is smaller than the one following from \eq{eq:mean-field1}. Also, the total time for the relaxation given by \eq{eq:mean-field2} is shorter. Increasing the initial velocity, the velocity oscillations develop, as shown in the inset of Fig.~\ref{dynamNEWGP}.

\section{Conclusions and discussions\label{sec:conclusions}}

We have studied the real-time nonequilibrium evolution of an impurity injected with a finite velocity into a system of weakly interacting bosons at zero temperature.  Our main goal was to gain an understanding of the dynamics in different regions of the parameter space, the relevant time scales and the final stationary state. We have shown that the dynamics exhibits a wealth of regimes. 

The relaxation proceeds in two main stages. First, an abrupt drop of the impurity velocity takes place due to emission of the dispersive density shock waves and the local hole formation. The impurity velocity reaches its minimal value in time at some characteristic relaxation time $T_r$. For not too heavy impurities with $M<M_c$,  we find that $T_r$ weakly depends on the initial impurity velocity. In contrast, it rapidely increases with the inverse dimensionless impurity coupling constant $\tilde{G}^{-1}$ and the ratio  $M/m$ of the impurity mass with respect to the mass of the bosons. This behaviour is illustrated in Figs.~\ref{fig2}, \ref{Voft9m}, \ref{fig:Gdependence} and \ref{figV2.5}. Nevertheless, the time for a total relaxation into a stationary state depends strongly on all three mentioned parameters.

After this first stage of relaxation, a heavy impurity with a sufficiently big initial velocity enters the transient dynamical regime characterized by the damped oscillations of the impurity velocity in time around its final stationary-state velocity, before its complete equilibration. We characterize this transient regime in Sec.~\ref{sec:flutter}.  The underlying mechanism are the damped oscillations of the local depletion cloud around the impurity position due to an effective attractive interaction between them, as well as inertial effects.

The dynamics of fast impurities displaying slowly decaying oscillations of the impurity velocity in a one-dimensional strongly interacting system of bosons was dubbed quantum flutter \cite{QFlutterNature,knap2014quantum}. It was claimed that the quantum flutter cannot be captured by a hydrodynamic theory and the Gross-Pitaevskii equation \cite{QFlutterNature}. Also, it was emphasized that the phenomenon requires the strong coupling regime \cite{QFlutterNature}. Nevertheless, we have shown that the description within the Gross-Pitaevskii equation reveals the impurity velocity oscillations in the system of weakly interacting bosons. Moreover, the oscillations are much more pronounced for a weakly coupled impurity. 

Note that the oscillations are also amplified by increasing the repulsion between the bosons, while keeping the other parameters fixed. Thus it would be interesting to verify if in the system of strongly interacting bosons the oscillations also become more  pronounced at weak impurity coupling. 

Another interesting question concerns the stationary state of the system. We find that the final impurity velocity remains almost constant in a wide range of experimentally relevant values of initial velocities exceeding a certain threshold, see Sec.~\ref{sec:crossover}.  The physical mechanism is the emission of dispersive density shock waves triggered by the sudden impurity injection. The density waves carry away the excess of the initial impurity momentum, while the impurity final velocity and the local depletion of the boson density remain constant.
On the other hand, a heavy impurity with the same initial momentum and the coupling strength as a light one, exhibits a very different relaxation dynamics. Heavy impurity continues to further deplete the boson density locally as the system momentum increases and in such a way gets rid of its momentum and slows down.
As a result, the regime of constant final impurity velocity is absent.

In the process of relaxation, a heavy impurity can emit both the density waves and solitons. We stress that the latter does not necessary lead to a change of the impurity direction of motion. Ref.~\cite{Will_2023} points out that the soliton emission and the change of the direction of motion require the strong impurity coupling and fast impurities.
We showed that both phenomena actually take place beyond the two restrictions, as illustrated in Fig.~\ref{fig:Single_soliton}.
We explain the underlying physics. Namely, for an impurity mass bigger than the critical one, at initial momenta situated in the vicinity of the termination points of metastable branches, the impurity gets rid of its energy by emitting solitons as it relaxes into the ground state, see Fig.~\ref{fig:aboveMc}. In the close vicinity of the termination points of metastable branches the impurity dynamics is considerably slowed down and the impurity changes the direction of motion. Otherwise, for $M<M_c$, heavy and initially fast impurities can also reverse their direction of motion, see Figs.~\ref{figV2.5} and \ref{Voft9m}. 

\section{Acknowledgments}

We are grateful to Z. Ristivojevic for providing the data for the Yang-Gaudin model from Ref.~\cite{Zoran_YangG} shown in Fig.~\ref{fig:comparison}.
This study has been partially supported through the EUR grant NanoX n° ANR-17-EURE-0009 in the framework of the ``Programme des Investissements d'Avenir".


\begin{thebibliography}{60}%
\makeatletter
\providecommand \@ifxundefined [1]{%
 \@ifx{#1\undefined}
}%
\providecommand \@ifnum [1]{%
 \ifnum #1\expandafter \@firstoftwo
 \else \expandafter \@secondoftwo
 \fi
}%
\providecommand \@ifx [1]{%
 \ifx #1\expandafter \@firstoftwo
 \else \expandafter \@secondoftwo
 \fi
}%
\providecommand \natexlab [1]{#1}%
\providecommand \enquote  [1]{``#1''}%
\providecommand \bibnamefont  [1]{#1}%
\providecommand \bibfnamefont [1]{#1}%
\providecommand \citenamefont [1]{#1}%
\providecommand \href@noop [0]{\@secondoftwo}%
\providecommand \href [0]{\begingroup \@sanitize@url \@href}%
\providecommand \@href[1]{\@@startlink{#1}\@@href}%
\providecommand \@@href[1]{\endgroup#1\@@endlink}%
\providecommand \@sanitize@url [0]{\catcode `\\12\catcode `\$12\catcode
  `\&12\catcode `\#12\catcode `\^12\catcode `\_12\catcode `\%12\relax}%
\providecommand \@@startlink[1]{}%
\providecommand \@@endlink[0]{}%
\providecommand \url  [0]{\begingroup\@sanitize@url \@url }%
\providecommand \@url [1]{\endgroup\@href {#1}{\urlprefix }}%
\providecommand \urlprefix  [0]{URL }%
\providecommand \Eprint [0]{\href }%
\providecommand \doibase [0]{http://dx.doi.org/}%
\providecommand \selectlanguage [0]{\@gobble}%
\providecommand \bibinfo  [0]{\@secondoftwo}%
\providecommand \bibfield  [0]{\@secondoftwo}%
\providecommand \translation [1]{[#1]}%
\providecommand \BibitemOpen [0]{}%
\providecommand \bibitemStop [0]{}%
\providecommand \bibitemNoStop [0]{.\EOS\space}%
\providecommand \EOS [0]{\spacefactor3000\relax}%
\providecommand \BibitemShut  [1]{\csname bibitem#1\endcsname}%
\let\auto@bib@innerbib\@empty
\bibitem [{\citenamefont {Zvonarev}\ \emph {et~al.}(2007)\citenamefont
  {Zvonarev}, \citenamefont {Cheianov},\ and\ \citenamefont
  {Giamarchi}}]{zvonarev2007spin}%
  \BibitemOpen
  \bibfield  {author} {\bibinfo {author} {\bibfnamefont {M.~B.}\ \bibnamefont
  {Zvonarev}}, \bibinfo {author} {\bibfnamefont {V.~V.}\ \bibnamefont
  {Cheianov}}, \ and\ \bibinfo {author} {\bibfnamefont {T.}~\bibnamefont
  {Giamarchi}},\ }\href {\doibase 10.1103/PhysRevLett.99.240404} {\bibfield
  {journal} {\bibinfo  {journal} {Phys. Rev. Lett.}\ }\textbf {\bibinfo
  {volume} {99}},\ \bibinfo {pages} {240404} (\bibinfo {year}
  {2007})}\BibitemShut {NoStop}%
\bibitem [{\citenamefont {Schecter}\ \emph
  {et~al.}(2012{\natexlab{a}})\citenamefont {Schecter}, \citenamefont
  {Gangardt},\ and\ \citenamefont {Kamenev}}]{AnnalsKamenev}%
  \BibitemOpen
  \bibfield  {author} {\bibinfo {author} {\bibfnamefont {M.}~\bibnamefont
  {Schecter}}, \bibinfo {author} {\bibfnamefont {D.}~\bibnamefont {Gangardt}},
  \ and\ \bibinfo {author} {\bibfnamefont {A.}~\bibnamefont {Kamenev}},\ }\href
  {\doibase https://doi.org/10.1016/j.aop.2011.10.001} {\bibfield  {journal}
  {\bibinfo  {journal} {Ann. Phys.}\ }\textbf {\bibinfo {volume} {327}},\
  \bibinfo {pages} {639} (\bibinfo {year} {2012}{\natexlab{a}})}\BibitemShut
  {NoStop}%
\bibitem [{\citenamefont {Peotta}\ \emph {et~al.}(2013)\citenamefont {Peotta},
  \citenamefont {Rossini}, \citenamefont {Polini}, \citenamefont {Minardi},\
  and\ \citenamefont {Fazio}}]{peotta2013quantum}%
  \BibitemOpen
  \bibfield  {author} {\bibinfo {author} {\bibfnamefont {S.}~\bibnamefont
  {Peotta}}, \bibinfo {author} {\bibfnamefont {D.}~\bibnamefont {Rossini}},
  \bibinfo {author} {\bibfnamefont {M.}~\bibnamefont {Polini}}, \bibinfo
  {author} {\bibfnamefont {F.}~\bibnamefont {Minardi}}, \ and\ \bibinfo
  {author} {\bibfnamefont {R.}~\bibnamefont {Fazio}},\ }\href {\doibase
  10.1103/PhysRevLett.110.015302} {\bibfield  {journal} {\bibinfo  {journal}
  {Phys. Rev. Lett.}\ }\textbf {\bibinfo {volume} {110}},\ \bibinfo {pages}
  {015302} (\bibinfo {year} {2013})}\BibitemShut {NoStop}%
\bibitem [{\citenamefont {Kantian}\ \emph {et~al.}(2014)\citenamefont
  {Kantian}, \citenamefont {Schollw\"ock},\ and\ \citenamefont
  {Giamarchi}}]{PhysRevLett.113.070601}%
  \BibitemOpen
  \bibfield  {author} {\bibinfo {author} {\bibfnamefont {A.}~\bibnamefont
  {Kantian}}, \bibinfo {author} {\bibfnamefont {U.}~\bibnamefont
  {Schollw\"ock}}, \ and\ \bibinfo {author} {\bibfnamefont {T.}~\bibnamefont
  {Giamarchi}},\ }\href {\doibase 10.1103/PhysRevLett.113.070601} {\bibfield
  {journal} {\bibinfo  {journal} {Phys. Rev. Lett.}\ }\textbf {\bibinfo
  {volume} {113}},\ \bibinfo {pages} {070601} (\bibinfo {year}
  {2014})}\BibitemShut {NoStop}%
\bibitem [{\citenamefont {Volosniev}\ \emph {et~al.}(2015)\citenamefont
  {Volosniev}, \citenamefont {Hammer},\ and\ \citenamefont
  {Zinner}}]{PhysRevA.92.023623}%
  \BibitemOpen
  \bibfield  {author} {\bibinfo {author} {\bibfnamefont {A.~G.}\ \bibnamefont
  {Volosniev}}, \bibinfo {author} {\bibfnamefont {H.-W.}\ \bibnamefont
  {Hammer}}, \ and\ \bibinfo {author} {\bibfnamefont {N.~T.}\ \bibnamefont
  {Zinner}},\ }\href {\doibase 10.1103/PhysRevA.92.023623} {\bibfield
  {journal} {\bibinfo  {journal} {Phys. Rev. A}\ }\textbf {\bibinfo {volume}
  {92}},\ \bibinfo {pages} {023623} (\bibinfo {year} {2015})}\BibitemShut
  {NoStop}%
\bibitem [{\citenamefont {Grusdt}\ \emph {et~al.}(2017)\citenamefont {Grusdt},
  \citenamefont {Astrakharchik},\ and\ \citenamefont {Demler}}]{Grusdt_2017}%
  \BibitemOpen
  \bibfield  {author} {\bibinfo {author} {\bibfnamefont {F.}~\bibnamefont
  {Grusdt}}, \bibinfo {author} {\bibfnamefont {G.~E.}\ \bibnamefont
  {Astrakharchik}}, \ and\ \bibinfo {author} {\bibfnamefont {E.}~\bibnamefont
  {Demler}},\ }\href {\doibase 10.1088/1367-2630/aa8a2e} {\bibfield  {journal}
  {\bibinfo  {journal} {New J. Phys.}\ }\textbf {\bibinfo {volume} {19}},\
  \bibinfo {pages} {103035} (\bibinfo {year} {2017})}\BibitemShut {NoStop}%
\bibitem [{\citenamefont {Lampo}\ \emph {et~al.}(2017)\citenamefont {Lampo},
  \citenamefont {Lim}, \citenamefont {Garc{\'{i}}a-March},\ and\ \citenamefont
  {Lewenstein}}]{LewensteinBrownien}%
  \BibitemOpen
  \bibfield  {author} {\bibinfo {author} {\bibfnamefont {A.}~\bibnamefont
  {Lampo}}, \bibinfo {author} {\bibfnamefont {S.~H.}\ \bibnamefont {Lim}},
  \bibinfo {author} {\bibfnamefont {M.~{\'{A}}.}\ \bibnamefont
  {Garc{\'{i}}a-March}}, \ and\ \bibinfo {author} {\bibfnamefont
  {M.}~\bibnamefont {Lewenstein}},\ }\href {\doibase 10.22331/q-2017-09-27-30}
  {\bibfield  {journal} {\bibinfo  {journal} {{Quantum}}\ }\textbf {\bibinfo
  {volume} {1}},\ \bibinfo {pages} {30} (\bibinfo {year} {2017})}\BibitemShut
  {NoStop}%
\bibitem [{\citenamefont {Mistakidis}\ \emph {et~al.}(2023)\citenamefont
  {Mistakidis}, \citenamefont {Volosniev}, \citenamefont {Barfknecht},
  \citenamefont {Fogarty}, \citenamefont {Busch}, \citenamefont {Foerster},
  \citenamefont {Schmelcher},\ and\ \citenamefont
  {Zinner}}]{PhysicsReports2023}%
  \BibitemOpen
  \bibfield  {author} {\bibinfo {author} {\bibfnamefont {S.}~\bibnamefont
  {Mistakidis}}, \bibinfo {author} {\bibfnamefont {A.}~\bibnamefont
  {Volosniev}}, \bibinfo {author} {\bibfnamefont {R.}~\bibnamefont
  {Barfknecht}}, \bibinfo {author} {\bibfnamefont {T.}~\bibnamefont {Fogarty}},
  \bibinfo {author} {\bibfnamefont {T.}~\bibnamefont {Busch}}, \bibinfo
  {author} {\bibfnamefont {A.}~\bibnamefont {Foerster}}, \bibinfo {author}
  {\bibfnamefont {P.}~\bibnamefont {Schmelcher}}, \ and\ \bibinfo {author}
  {\bibfnamefont {N.}~\bibnamefont {Zinner}},\ }\href {\doibase
  https://doi.org/10.1016/j.physrep.2023.10.004} {\bibfield  {journal}
  {\bibinfo  {journal} {Physics Reports}\ }\textbf {\bibinfo {volume} {1042}},\
  \bibinfo {pages} {1} (\bibinfo {year} {2023})}\BibitemShut {NoStop}%
\bibitem [{\citenamefont {Gamayun}\ and\ \citenamefont
  {Lychkovskiy}(2024)}]{Gamayun2024}%
  \BibitemOpen
  \bibfield  {author} {\bibinfo {author} {\bibfnamefont {O.}~\bibnamefont
  {Gamayun}}\ and\ \bibinfo {author} {\bibfnamefont {O.}~\bibnamefont
  {Lychkovskiy}},\ }\href {\doibase 10.21468/SciPostPhys.17.2.063} {\bibfield
  {journal} {\bibinfo  {journal} {SciPost Phys.}\ }\textbf {\bibinfo {volume}
  {17}},\ \bibinfo {pages} {063} (\bibinfo {year} {2024})}\BibitemShut
  {NoStop}%
\bibitem [{\citenamefont {Gangardt}\ and\ \citenamefont
  {Kamenev}(2009)}]{gangardt2009bloch}%
  \BibitemOpen
  \bibfield  {author} {\bibinfo {author} {\bibfnamefont {D.~M.}\ \bibnamefont
  {Gangardt}}\ and\ \bibinfo {author} {\bibfnamefont {A.}~\bibnamefont
  {Kamenev}},\ }\href {\doibase 10.1103/PhysRevLett.102.070402} {\bibfield
  {journal} {\bibinfo  {journal} {Phys. Rev. Lett.}\ }\textbf {\bibinfo
  {volume} {102}},\ \bibinfo {pages} {070402} (\bibinfo {year}
  {2009})}\BibitemShut {NoStop}%
\bibitem [{\citenamefont {Meinert}\ \emph {et~al.}(2017)\citenamefont
  {Meinert}, \citenamefont {Knap}, \citenamefont {Kirilov}, \citenamefont
  {Jag-Lauber}, \citenamefont {Zvonarev}, \citenamefont {Demler},\ and\
  \citenamefont {N{\"a}gerl}}]{Meinert945}%
  \BibitemOpen
  \bibfield  {author} {\bibinfo {author} {\bibfnamefont {F.}~\bibnamefont
  {Meinert}}, \bibinfo {author} {\bibfnamefont {M.}~\bibnamefont {Knap}},
  \bibinfo {author} {\bibfnamefont {E.}~\bibnamefont {Kirilov}}, \bibinfo
  {author} {\bibfnamefont {K.}~\bibnamefont {Jag-Lauber}}, \bibinfo {author}
  {\bibfnamefont {M.~B.}\ \bibnamefont {Zvonarev}}, \bibinfo {author}
  {\bibfnamefont {E.}~\bibnamefont {Demler}}, \ and\ \bibinfo {author}
  {\bibfnamefont {H.-C.}\ \bibnamefont {N{\"a}gerl}},\ }\href {\doibase
  10.1126/science.aah6616} {\bibfield  {journal} {\bibinfo  {journal}
  {Science}\ }\textbf {\bibinfo {volume} {356}},\ \bibinfo {pages} {945}
  (\bibinfo {year} {2017})}\BibitemShut {NoStop}%
\bibitem [{\citenamefont {Lamacraft}(2009)}]{lamacraft2009dispersion}%
  \BibitemOpen
  \bibfield  {author} {\bibinfo {author} {\bibfnamefont {A.}~\bibnamefont
  {Lamacraft}},\ }\href {\doibase 10.1103/PhysRevB.79.241105} {\bibfield
  {journal} {\bibinfo  {journal} {Phys. Rev. B}\ }\textbf {\bibinfo {volume}
  {79}},\ \bibinfo {pages} {241105} (\bibinfo {year} {2009})}\BibitemShut
  {NoStop}%
\bibitem [{\citenamefont {Kamenev}\ and\ \citenamefont
  {Glazman}(2009)}]{kamenev2009dynamics}%
  \BibitemOpen
  \bibfield  {author} {\bibinfo {author} {\bibfnamefont {A.}~\bibnamefont
  {Kamenev}}\ and\ \bibinfo {author} {\bibfnamefont {L.~I.}\ \bibnamefont
  {Glazman}},\ }\href {\doibase 10.1103/PhysRevA.80.011603} {\bibfield
  {journal} {\bibinfo  {journal} {Phys. Rev. A}\ }\textbf {\bibinfo {volume}
  {80}},\ \bibinfo {pages} {011603} (\bibinfo {year} {2009})}\BibitemShut
  {NoStop}%
\bibitem [{\citenamefont {Landau}\ and\ \citenamefont
  {Lifshitz}(1987)}]{Landau}%
  \BibitemOpen
  \bibfield  {author} {\bibinfo {author} {\bibfnamefont {L.~D.}\ \bibnamefont
  {Landau}}\ and\ \bibinfo {author} {\bibfnamefont {E.~M.}\ \bibnamefont
  {Lifshitz}},\ }\href@noop {} {\emph {\bibinfo {title} {Electrodynamics of
  Continuous Media}}}\ (\bibinfo  {publisher} {Pergamon Press},\ \bibinfo
  {year} {1987})\BibitemShut {NoStop}%
\bibitem [{\citenamefont {Landau}\ and\ \citenamefont
  {Khalatnikov}(1949)}]{landau+49}%
  \BibitemOpen
  \bibfield  {author} {\bibinfo {author} {\bibfnamefont {L.~D.}\ \bibnamefont
  {Landau}}\ and\ \bibinfo {author} {\bibfnamefont {I.~M.}\ \bibnamefont
  {Khalatnikov}},\ }\href@noop {} {\bibfield  {journal} {\bibinfo  {journal}
  {Zh. Eksp. Teor. Fiz.}\ }\textbf {\bibinfo {volume} {19}},\ \bibinfo {pages}
  {709} (\bibinfo {year} {1949})}\BibitemShut {NoStop}%
\bibitem [{\citenamefont {Hakim}(1997)}]{Hakim}%
  \BibitemOpen
  \bibfield  {author} {\bibinfo {author} {\bibfnamefont {V.}~\bibnamefont
  {Hakim}},\ }\href {\doibase 10.1103/PhysRevE.55.2835} {\bibfield  {journal}
  {\bibinfo  {journal} {Phys. Rev. E}\ }\textbf {\bibinfo {volume} {55}},\
  \bibinfo {pages} {2835} (\bibinfo {year} {1997})}\BibitemShut {NoStop}%
\bibitem [{\citenamefont {Pavloff}(2002)}]{PhysRevA.66.013610}%
  \BibitemOpen
  \bibfield  {author} {\bibinfo {author} {\bibfnamefont {N.}~\bibnamefont
  {Pavloff}},\ }\href {\doibase 10.1103/PhysRevA.66.013610} {\bibfield
  {journal} {\bibinfo  {journal} {Phys. Rev. A}\ }\textbf {\bibinfo {volume}
  {66}},\ \bibinfo {pages} {013610} (\bibinfo {year} {2002})}\BibitemShut
  {NoStop}%
\bibitem [{\citenamefont {Astrakharchik}\ and\ \citenamefont
  {Pitaevskii}(2004)}]{astrakharchik2004motion}%
  \BibitemOpen
  \bibfield  {author} {\bibinfo {author} {\bibfnamefont {G.~E.}\ \bibnamefont
  {Astrakharchik}}\ and\ \bibinfo {author} {\bibfnamefont {L.~P.}\ \bibnamefont
  {Pitaevskii}},\ }\href {\doibase 10.1103/PhysRevA.70.013608} {\bibfield
  {journal} {\bibinfo  {journal} {Phys. Rev. A}\ }\textbf {\bibinfo {volume}
  {70}},\ \bibinfo {pages} {013608} (\bibinfo {year} {2004})}\BibitemShut
  {NoStop}%
\bibitem [{\citenamefont {Carusotto}\ \emph {et~al.}(2006)\citenamefont
  {Carusotto}, \citenamefont {Hu}, \citenamefont {Collins},\ and\ \citenamefont
  {Smerzi}}]{PRLCarusotto2006}%
  \BibitemOpen
  \bibfield  {author} {\bibinfo {author} {\bibfnamefont {I.}~\bibnamefont
  {Carusotto}}, \bibinfo {author} {\bibfnamefont {S.~X.}\ \bibnamefont {Hu}},
  \bibinfo {author} {\bibfnamefont {L.~A.}\ \bibnamefont {Collins}}, \ and\
  \bibinfo {author} {\bibfnamefont {A.}~\bibnamefont {Smerzi}},\ }\href
  {\doibase 10.1103/PhysRevLett.97.260403} {\bibfield  {journal} {\bibinfo
  {journal} {Phys. Rev. Lett.}\ }\textbf {\bibinfo {volume} {97}},\ \bibinfo
  {pages} {260403} (\bibinfo {year} {2006})}\BibitemShut {NoStop}%
\bibitem [{\citenamefont {Cherny}\ \emph {et~al.}(2012)\citenamefont {Cherny},
  \citenamefont {Caux},\ and\ \citenamefont {Brand}}]{Cherny2012}%
  \BibitemOpen
  \bibfield  {author} {\bibinfo {author} {\bibfnamefont {A.~Y.}\ \bibnamefont
  {Cherny}}, \bibinfo {author} {\bibfnamefont {J.-S.}\ \bibnamefont {Caux}}, \
  and\ \bibinfo {author} {\bibfnamefont {J.}~\bibnamefont {Brand}},\ }\href
  {\doibase 10.1007/s11467-011-0211-2} {\bibfield  {journal} {\bibinfo
  {journal} {Front. Phys.}\ }\textbf {\bibinfo {volume} {7}},\ \bibinfo {pages}
  {54} (\bibinfo {year} {2012})}\BibitemShut {NoStop}%
\bibitem [{\citenamefont {Mathy}\ \emph {et~al.}(2012)\citenamefont {Mathy},
  \citenamefont {Zvonarev},\ and\ \citenamefont {Demler}}]{QFlutterNature}%
  \BibitemOpen
  \bibfield  {author} {\bibinfo {author} {\bibfnamefont {C.}~\bibnamefont
  {Mathy}}, \bibinfo {author} {\bibfnamefont {M.}~\bibnamefont {Zvonarev}}, \
  and\ \bibinfo {author} {\bibfnamefont {E.}~\bibnamefont {Demler}},\ }\href
  {\doibase https://doi.org/10.1038/nphys2455} {\bibfield  {journal} {\bibinfo
  {journal} {Nature Phys.}\ }\textbf {\bibinfo {volume} {8}},\ \bibinfo {pages}
  {881} (\bibinfo {year} {2012})}\BibitemShut {NoStop}%
\bibitem [{\citenamefont {Gamayun}\ \emph {et~al.}(2014)\citenamefont
  {Gamayun}, \citenamefont {Lychkovskiy},\ and\ \citenamefont
  {Cheianov}}]{PhysRevE.90.032132}%
  \BibitemOpen
  \bibfield  {author} {\bibinfo {author} {\bibfnamefont {O.}~\bibnamefont
  {Gamayun}}, \bibinfo {author} {\bibfnamefont {O.}~\bibnamefont
  {Lychkovskiy}}, \ and\ \bibinfo {author} {\bibfnamefont {V.}~\bibnamefont
  {Cheianov}},\ }\href {\doibase 10.1103/PhysRevE.90.032132} {\bibfield
  {journal} {\bibinfo  {journal} {Phys. Rev. E}\ }\textbf {\bibinfo {volume}
  {90}},\ \bibinfo {pages} {032132} (\bibinfo {year} {2014})}\BibitemShut
  {NoStop}%
\bibitem [{\citenamefont {Knap}\ \emph {et~al.}(2014)\citenamefont {Knap},
  \citenamefont {Mathy}, \citenamefont {Ganahl}, \citenamefont {Zvonarev},\
  and\ \citenamefont {Demler}}]{knap2014quantum}%
  \BibitemOpen
  \bibfield  {author} {\bibinfo {author} {\bibfnamefont {M.}~\bibnamefont
  {Knap}}, \bibinfo {author} {\bibfnamefont {C.~J.~M.}\ \bibnamefont {Mathy}},
  \bibinfo {author} {\bibfnamefont {M.}~\bibnamefont {Ganahl}}, \bibinfo
  {author} {\bibfnamefont {M.~B.}\ \bibnamefont {Zvonarev}}, \ and\ \bibinfo
  {author} {\bibfnamefont {E.}~\bibnamefont {Demler}},\ }\href {\doibase
  10.1103/PhysRevLett.112.015302} {\bibfield  {journal} {\bibinfo  {journal}
  {Phys. Rev. Lett.}\ }\textbf {\bibinfo {volume} {112}},\ \bibinfo {pages}
  {015302} (\bibinfo {year} {2014})}\BibitemShut {NoStop}%
\bibitem [{\citenamefont {Robinson}\ \emph {et~al.}(2016)\citenamefont
  {Robinson}, \citenamefont {Caux},\ and\ \citenamefont
  {Konik}}]{robinson2016motion}%
  \BibitemOpen
  \bibfield  {author} {\bibinfo {author} {\bibfnamefont {N.~J.}\ \bibnamefont
  {Robinson}}, \bibinfo {author} {\bibfnamefont {J.-S.}\ \bibnamefont {Caux}},
  \ and\ \bibinfo {author} {\bibfnamefont {R.~M.}\ \bibnamefont {Konik}},\
  }\href {\doibase 10.1103/PhysRevLett.116.145302} {\bibfield  {journal}
  {\bibinfo  {journal} {Phys. Rev. Lett.}\ }\textbf {\bibinfo {volume} {116}},\
  \bibinfo {pages} {145302} (\bibinfo {year} {2016})}\BibitemShut {NoStop}%
\bibitem [{\citenamefont {Gamayun}\ \emph {et~al.}(2018)\citenamefont
  {Gamayun}, \citenamefont {Lychkovskiy}, \citenamefont {Burovski},
  \citenamefont {Malcomson}, \citenamefont {Cheianov},\ and\ \citenamefont
  {Zvonarev}}]{QuenchZvonarev}%
  \BibitemOpen
  \bibfield  {author} {\bibinfo {author} {\bibfnamefont {O.}~\bibnamefont
  {Gamayun}}, \bibinfo {author} {\bibfnamefont {O.}~\bibnamefont
  {Lychkovskiy}}, \bibinfo {author} {\bibfnamefont {E.}~\bibnamefont
  {Burovski}}, \bibinfo {author} {\bibfnamefont {M.}~\bibnamefont {Malcomson}},
  \bibinfo {author} {\bibfnamefont {V.~V.}\ \bibnamefont {Cheianov}}, \ and\
  \bibinfo {author} {\bibfnamefont {M.~B.}\ \bibnamefont {Zvonarev}},\ }\href
  {\doibase 10.1103/PhysRevLett.120.220605} {\bibfield  {journal} {\bibinfo
  {journal} {Phys. Rev. Lett.}\ }\textbf {\bibinfo {volume} {120}},\ \bibinfo
  {pages} {220605} (\bibinfo {year} {2018})}\BibitemShut {NoStop}%
\bibitem [{\citenamefont {Koutentakis}\ \emph {et~al.}(2022)\citenamefont
  {Koutentakis}, \citenamefont {Mistakidis},\ and\ \citenamefont
  {Schmelcher}}]{atoms10010003}%
  \BibitemOpen
  \bibfield  {author} {\bibinfo {author} {\bibfnamefont {G.~M.}\ \bibnamefont
  {Koutentakis}}, \bibinfo {author} {\bibfnamefont {S.~I.}\ \bibnamefont
  {Mistakidis}}, \ and\ \bibinfo {author} {\bibfnamefont {P.}~\bibnamefont
  {Schmelcher}},\ }\href {\doibase 10.3390/atoms10010003} {\bibfield  {journal}
  {\bibinfo  {journal} {Atoms}\ }\textbf {\bibinfo {volume} {10}} (\bibinfo
  {year} {2022}),\ 10.3390/atoms10010003}\BibitemShut {NoStop}%
\bibitem [{\citenamefont {Will}\ and\ \citenamefont
  {Fleischhauer}(2023)}]{Will_2023}%
  \BibitemOpen
  \bibfield  {author} {\bibinfo {author} {\bibfnamefont {M.}~\bibnamefont
  {Will}}\ and\ \bibinfo {author} {\bibfnamefont {M.}~\bibnamefont
  {Fleischhauer}},\ }\href {\doibase 10.1088/1367-2630/acf06a} {\bibfield
  {journal} {\bibinfo  {journal} {New J. of Phys.}\ }\textbf {\bibinfo {volume}
  {25}},\ \bibinfo {pages} {083043} (\bibinfo {year} {2023})}\BibitemShut
  {NoStop}%
\bibitem [{\citenamefont {Zhang}\ \emph {et~al.}(2024)\citenamefont {Zhang},
  \citenamefont {Jiang}, \citenamefont {Lin},\ and\ \citenamefont
  {Guan}}]{YangGaudinFlutter}%
  \BibitemOpen
  \bibfield  {author} {\bibinfo {author} {\bibfnamefont {Z.-H.}\ \bibnamefont
  {Zhang}}, \bibinfo {author} {\bibfnamefont {Y.}~\bibnamefont {Jiang}},
  \bibinfo {author} {\bibfnamefont {H.-Q.}\ \bibnamefont {Lin}}, \ and\
  \bibinfo {author} {\bibfnamefont {X.-W.}\ \bibnamefont {Guan}},\ }\href
  {\doibase 10.1103/PhysRevA.110.023329} {\bibfield  {journal} {\bibinfo
  {journal} {Phys. Rev. A}\ }\textbf {\bibinfo {volume} {110}},\ \bibinfo
  {pages} {023329} (\bibinfo {year} {2024})}\BibitemShut {NoStop}%
\bibitem [{\citenamefont {Grusdt}\ \emph {et~al.}(2025)\citenamefont {Grusdt},
  \citenamefont {Mostaan}, \citenamefont {Demler},\ and\ \citenamefont
  {Ardila}}]{grusdt2025impurities}%
  \BibitemOpen
  \bibfield  {author} {\bibinfo {author} {\bibfnamefont {F.}~\bibnamefont
  {Grusdt}}, \bibinfo {author} {\bibfnamefont {N.}~\bibnamefont {Mostaan}},
  \bibinfo {author} {\bibfnamefont {E.}~\bibnamefont {Demler}}, \ and\ \bibinfo
  {author} {\bibfnamefont {L.~A. A. P.~P.}\ \bibnamefont {Ardila}},\
  }\href@noop {} {\bibfield  {journal} {\bibinfo  {journal} {Reports on
  Progress in Physics}\ } (\bibinfo {year} {2025})}\BibitemShut {NoStop}%
\bibitem [{\citenamefont {Palzer}\ \emph {et~al.}(2009)\citenamefont {Palzer},
  \citenamefont {Zipkes}, \citenamefont {Sias},\ and\ \citenamefont
  {K{\"o}hl}}]{palzer2009quantum}%
  \BibitemOpen
  \bibfield  {author} {\bibinfo {author} {\bibfnamefont {S.}~\bibnamefont
  {Palzer}}, \bibinfo {author} {\bibfnamefont {C.}~\bibnamefont {Zipkes}},
  \bibinfo {author} {\bibfnamefont {C.}~\bibnamefont {Sias}}, \ and\ \bibinfo
  {author} {\bibfnamefont {M.}~\bibnamefont {K{\"o}hl}},\ }\href {\doibase
  10.1103/PhysRevLett.103.150601} {\bibfield  {journal} {\bibinfo  {journal}
  {Phys. Rev. Lett.}\ }\textbf {\bibinfo {volume} {103}},\ \bibinfo {pages}
  {150601} (\bibinfo {year} {2009})}\BibitemShut {NoStop}%
\bibitem [{\citenamefont {Catani}\ \emph {et~al.}(2012)\citenamefont {Catani},
  \citenamefont {Lamporesi}, \citenamefont {Naik}, \citenamefont {Gring},
  \citenamefont {Inguscio}, \citenamefont {Minardi}, \citenamefont {Kantian},\
  and\ \citenamefont {Giamarchi}}]{2012quantum}%
  \BibitemOpen
  \bibfield  {author} {\bibinfo {author} {\bibfnamefont {J.}~\bibnamefont
  {Catani}}, \bibinfo {author} {\bibfnamefont {G.}~\bibnamefont {Lamporesi}},
  \bibinfo {author} {\bibfnamefont {D.}~\bibnamefont {Naik}}, \bibinfo {author}
  {\bibfnamefont {M.}~\bibnamefont {Gring}}, \bibinfo {author} {\bibfnamefont
  {M.}~\bibnamefont {Inguscio}}, \bibinfo {author} {\bibfnamefont
  {F.}~\bibnamefont {Minardi}}, \bibinfo {author} {\bibfnamefont
  {A.}~\bibnamefont {Kantian}}, \ and\ \bibinfo {author} {\bibfnamefont
  {T.}~\bibnamefont {Giamarchi}},\ }\href {\doibase 10.1103/PhysRevA.85.023623}
  {\bibfield  {journal} {\bibinfo  {journal} {Phys. Rev. A}\ }\textbf {\bibinfo
  {volume} {85}},\ \bibinfo {pages} {023623} (\bibinfo {year}
  {2012})}\BibitemShut {NoStop}%
\bibitem [{\citenamefont {Fukuhara}\ \emph {et~al.}(2013)\citenamefont
  {Fukuhara}, \citenamefont {Kantian}, \citenamefont {Endres}, \citenamefont
  {Cheneau}, \citenamefont {Schau{\ss}}, \citenamefont {Hild}, \citenamefont
  {Bellem}, \citenamefont {Schollw{\"o}ck}, \citenamefont {Giamarchi},
  \citenamefont {Gross}, \citenamefont {Bloch},\ and\ \citenamefont
  {Kuhr}}]{fukuhara2013quantum}%
  \BibitemOpen
  \bibfield  {author} {\bibinfo {author} {\bibfnamefont {T.}~\bibnamefont
  {Fukuhara}}, \bibinfo {author} {\bibfnamefont {A.}~\bibnamefont {Kantian}},
  \bibinfo {author} {\bibfnamefont {M.}~\bibnamefont {Endres}}, \bibinfo
  {author} {\bibfnamefont {M.}~\bibnamefont {Cheneau}}, \bibinfo {author}
  {\bibfnamefont {P.}~\bibnamefont {Schau{\ss}}}, \bibinfo {author}
  {\bibfnamefont {S.}~\bibnamefont {Hild}}, \bibinfo {author} {\bibfnamefont
  {D.}~\bibnamefont {Bellem}}, \bibinfo {author} {\bibfnamefont
  {U.}~\bibnamefont {Schollw{\"o}ck}}, \bibinfo {author} {\bibfnamefont
  {T.}~\bibnamefont {Giamarchi}}, \bibinfo {author} {\bibfnamefont
  {C.}~\bibnamefont {Gross}}, \bibinfo {author} {\bibfnamefont
  {I.}~\bibnamefont {Bloch}}, \ and\ \bibinfo {author} {\bibfnamefont
  {S.}~\bibnamefont {Kuhr}},\ }\href {\doibase 10.1038/nphys2561} {\bibfield
  {journal} {\bibinfo  {journal} {Nat. Phys.}\ }\textbf {\bibinfo {volume}
  {9}},\ \bibinfo {pages} {235} (\bibinfo {year} {2013})}\BibitemShut {NoStop}%
\bibitem [{\citenamefont {Lee}\ \emph {et~al.}(1953)\citenamefont {Lee},
  \citenamefont {Low},\ and\ \citenamefont {Pines}}]{LeeLowPines}%
  \BibitemOpen
  \bibfield  {author} {\bibinfo {author} {\bibfnamefont {T.~D.}\ \bibnamefont
  {Lee}}, \bibinfo {author} {\bibfnamefont {F.~E.}\ \bibnamefont {Low}}, \ and\
  \bibinfo {author} {\bibfnamefont {D.}~\bibnamefont {Pines}},\ }\href
  {\doibase https://doi.org/10.1103/PhysRev.90.297} {\bibfield  {journal}
  {\bibinfo  {journal} {Phys. Rev.}\ }\textbf {\bibinfo {volume} {90}},\
  \bibinfo {pages} {297} (\bibinfo {year} {1953})}\BibitemShut {NoStop}%
\bibitem [{\citenamefont {Pitaevskii}\ and\ \citenamefont
  {Stringari}(2003)}]{pitaevskii_bose-einstein_2003}%
  \BibitemOpen
  \bibfield  {author} {\bibinfo {author} {\bibfnamefont {L.~P.}\ \bibnamefont
  {Pitaevskii}}\ and\ \bibinfo {author} {\bibfnamefont {S.}~\bibnamefont
  {Stringari}},\ }\href@noop {} {\emph {\bibinfo {title} {Bose-{Einstein}
  {Condensation}}}},\ International {Series} of {Monographs} on {Physics}\
  (\bibinfo  {publisher} {Oxford University Press},\ \bibinfo {address}
  {Oxford, New York},\ \bibinfo {year} {2003})\BibitemShut {NoStop}%
\bibitem [{\citenamefont {Sykes}\ \emph {et~al.}(2009)\citenamefont {Sykes},
  \citenamefont {Davis},\ and\ \citenamefont {Roberts}}]{sykes_drag_2009}%
  \BibitemOpen
  \bibfield  {author} {\bibinfo {author} {\bibfnamefont {A.~G.}\ \bibnamefont
  {Sykes}}, \bibinfo {author} {\bibfnamefont {M.~J.}\ \bibnamefont {Davis}}, \
  and\ \bibinfo {author} {\bibfnamefont {D.~C.}\ \bibnamefont {Roberts}},\
  }\href {\doibase 10.1103/PhysRevLett.103.085302} {\bibfield  {journal}
  {\bibinfo  {journal} {Phys. Rev. Lett.}\ }\textbf {\bibinfo {volume} {103}},\
  \bibinfo {pages} {085302} (\bibinfo {year} {2009})}\BibitemShut {NoStop}%
\bibitem [{\citenamefont {Reichert}\ \emph {et~al.}(2019)\citenamefont
  {Reichert}, \citenamefont {Ristivojevic},\ and\ \citenamefont
  {Petkovi{\'{c}}}}]{CasimirNewJPhys}%
  \BibitemOpen
  \bibfield  {author} {\bibinfo {author} {\bibfnamefont {B.}~\bibnamefont
  {Reichert}}, \bibinfo {author} {\bibfnamefont {Z.}~\bibnamefont
  {Ristivojevic}}, \ and\ \bibinfo {author} {\bibfnamefont {A.}~\bibnamefont
  {Petkovi{\'{c}}}},\ }\href {\doibase 10.1088/1367-2630/ab1b8e} {\bibfield
  {journal} {\bibinfo  {journal} {New J. Phys.}\ }\textbf {\bibinfo {volume}
  {21}},\ \bibinfo {pages} {053024} (\bibinfo {year} {2019})}\BibitemShut
  {NoStop}%
\bibitem [{\citenamefont {Volosniev}\ and\ \citenamefont
  {Hammer}(2017)}]{VolosnievPolaron}%
  \BibitemOpen
  \bibfield  {author} {\bibinfo {author} {\bibfnamefont {A.~G.}\ \bibnamefont
  {Volosniev}}\ and\ \bibinfo {author} {\bibfnamefont {H.-W.}\ \bibnamefont
  {Hammer}},\ }\href {\doibase 10.1103/PhysRevA.96.031601} {\bibfield
  {journal} {\bibinfo  {journal} {Phys. Rev. A}\ }\textbf {\bibinfo {volume}
  {96}},\ \bibinfo {pages} {031601} (\bibinfo {year} {2017})}\BibitemShut
  {NoStop}%
\bibitem [{\citenamefont {Mistakidis}\ \emph {et~al.}(2019)\citenamefont
  {Mistakidis}, \citenamefont {Volosniev}, \citenamefont {Zinner},\ and\
  \citenamefont {Schmelcher}}]{PhysRevA.100.013619}%
  \BibitemOpen
  \bibfield  {author} {\bibinfo {author} {\bibfnamefont {S.~I.}\ \bibnamefont
  {Mistakidis}}, \bibinfo {author} {\bibfnamefont {A.~G.}\ \bibnamefont
  {Volosniev}}, \bibinfo {author} {\bibfnamefont {N.~T.}\ \bibnamefont
  {Zinner}}, \ and\ \bibinfo {author} {\bibfnamefont {P.}~\bibnamefont
  {Schmelcher}},\ }\href {\doibase 10.1103/PhysRevA.100.013619} {\bibfield
  {journal} {\bibinfo  {journal} {Phys. Rev. A}\ }\textbf {\bibinfo {volume}
  {100}},\ \bibinfo {pages} {013619} (\bibinfo {year} {2019})}\BibitemShut
  {NoStop}%
\bibitem [{\citenamefont {Tsuzuki}(1971)}]{tsuzuki_nonlinear_1971}%
  \BibitemOpen
  \bibfield  {author} {\bibinfo {author} {\bibfnamefont {T.}~\bibnamefont
  {Tsuzuki}},\ }\href {https://doi.org/10.1007/BF00628744} {\bibfield
  {journal} {\bibinfo  {journal} {J. Low Temp. Phys.}\ }\textbf {\bibinfo
  {volume} {4}},\ \bibinfo {pages} {441} (\bibinfo {year} {1971})}\BibitemShut
  {NoStop}%
\bibitem [{\citenamefont {Olshanii}(1998)}]{olshanii_1998}%
  \BibitemOpen
  \bibfield  {author} {\bibinfo {author} {\bibfnamefont {M.}~\bibnamefont
  {Olshanii}},\ }\href {\doibase 10.1103/PhysRevLett.81.938} {\bibfield
  {journal} {\bibinfo  {journal} {Phys. Rev. Lett.}\ }\textbf {\bibinfo
  {volume} {81}},\ \bibinfo {pages} {938} (\bibinfo {year} {1998})}\BibitemShut
  {NoStop}%
\bibitem [{\citenamefont {Peano}\ \emph {et~al.}(2005)\citenamefont {Peano},
  \citenamefont {Thorwart}, \citenamefont {Mora},\ and\ \citenamefont
  {Egger}}]{Peano_2005}%
  \BibitemOpen
  \bibfield  {author} {\bibinfo {author} {\bibfnamefont {V.}~\bibnamefont
  {Peano}}, \bibinfo {author} {\bibfnamefont {M.}~\bibnamefont {Thorwart}},
  \bibinfo {author} {\bibfnamefont {C.}~\bibnamefont {Mora}}, \ and\ \bibinfo
  {author} {\bibfnamefont {R.}~\bibnamefont {Egger}},\ }\href {\doibase
  10.1088/1367-2630/7/1/192} {\bibfield  {journal} {\bibinfo  {journal} {New
  Journal of Physics}\ }\textbf {\bibinfo {volume} {7}},\ \bibinfo {pages}
  {192} (\bibinfo {year} {2005})}\BibitemShut {NoStop}%
\bibitem [{\citenamefont {Bloch}\ \emph {et~al.}(2008)\citenamefont {Bloch},
  \citenamefont {Dalibard},\ and\ \citenamefont {Zwerger}}]{RevModPhys.80.885}%
  \BibitemOpen
  \bibfield  {author} {\bibinfo {author} {\bibfnamefont {I.}~\bibnamefont
  {Bloch}}, \bibinfo {author} {\bibfnamefont {J.}~\bibnamefont {Dalibard}}, \
  and\ \bibinfo {author} {\bibfnamefont {W.}~\bibnamefont {Zwerger}},\ }\href
  {\doibase 10.1103/RevModPhys.80.885} {\bibfield  {journal} {\bibinfo
  {journal} {Rev. Mod. Phys.}\ }\textbf {\bibinfo {volume} {80}},\ \bibinfo
  {pages} {885} (\bibinfo {year} {2008})}\BibitemShut {NoStop}%
\bibitem [{\citenamefont {Jager}\ \emph {et~al.}(2020)\citenamefont {Jager},
  \citenamefont {Barnett}, \citenamefont {Will},\ and\ \citenamefont
  {Fleischhauer}}]{PhysRevResearch.2.033142}%
  \BibitemOpen
  \bibfield  {author} {\bibinfo {author} {\bibfnamefont {J.}~\bibnamefont
  {Jager}}, \bibinfo {author} {\bibfnamefont {R.}~\bibnamefont {Barnett}},
  \bibinfo {author} {\bibfnamefont {M.}~\bibnamefont {Will}}, \ and\ \bibinfo
  {author} {\bibfnamefont {M.}~\bibnamefont {Fleischhauer}},\ }\href {\doibase
  10.1103/PhysRevResearch.2.033142} {\bibfield  {journal} {\bibinfo  {journal}
  {Phys. Rev. Res.}\ }\textbf {\bibinfo {volume} {2}},\ \bibinfo {pages}
  {033142} (\bibinfo {year} {2020})}\BibitemShut {NoStop}%
\bibitem [{\citenamefont {Panochko}\ and\ \citenamefont
  {Pastukhov}(2019)}]{Pastukhov2019}%
  \BibitemOpen
  \bibfield  {author} {\bibinfo {author} {\bibfnamefont {G.}~\bibnamefont
  {Panochko}}\ and\ \bibinfo {author} {\bibfnamefont {V.}~\bibnamefont
  {Pastukhov}},\ }\href {\doibase https://doi.org/10.1016/j.aop.2019.167933}
  {\bibfield  {journal} {\bibinfo  {journal} {Ann. Phys.}\ }\textbf {\bibinfo
  {volume} {409}},\ \bibinfo {pages} {167933} (\bibinfo {year}
  {2019})}\BibitemShut {NoStop}%
\bibitem [{\citenamefont {Petkovi\ifmmode~\acute{c}\else \'{c}\fi{}}\ and\
  \citenamefont {Ristivojevic}(2022)}]{PetkovicPRAL}%
  \BibitemOpen
  \bibfield  {author} {\bibinfo {author} {\bibfnamefont {A.}~\bibnamefont
  {Petkovi\ifmmode~\acute{c}\else \'{c}\fi{}}}\ and\ \bibinfo {author}
  {\bibfnamefont {Z.}~\bibnamefont {Ristivojevic}},\ }\href {\doibase
  10.1103/PhysRevA.105.L021303} {\bibfield  {journal} {\bibinfo  {journal}
  {Phys. Rev. A}\ }\textbf {\bibinfo {volume} {105}},\ \bibinfo {pages}
  {L021303} (\bibinfo {year} {2022})}\BibitemShut {NoStop}%
\bibitem [{\citenamefont {Parisi}\ and\ \citenamefont
  {Giorgini}(2017)}]{PhysRevA.95.023619}%
  \BibitemOpen
  \bibfield  {author} {\bibinfo {author} {\bibfnamefont {L.}~\bibnamefont
  {Parisi}}\ and\ \bibinfo {author} {\bibfnamefont {S.}~\bibnamefont
  {Giorgini}},\ }\href {\doibase 10.1103/PhysRevA.95.023619} {\bibfield
  {journal} {\bibinfo  {journal} {Phys. Rev. A}\ }\textbf {\bibinfo {volume}
  {95}},\ \bibinfo {pages} {023619} (\bibinfo {year} {2017})}\BibitemShut
  {NoStop}%
\bibitem [{\citenamefont {Yang}\ \emph {et~al.}(2022)\citenamefont {Yang},
  \citenamefont {Čufar}, \citenamefont {Pahl},\ and\ \citenamefont
  {Brand}}]{condmat7010015}%
  \BibitemOpen
  \bibfield  {author} {\bibinfo {author} {\bibfnamefont {M.}~\bibnamefont
  {Yang}}, \bibinfo {author} {\bibfnamefont {M.}~\bibnamefont {Čufar}},
  \bibinfo {author} {\bibfnamefont {E.}~\bibnamefont {Pahl}}, \ and\ \bibinfo
  {author} {\bibfnamefont {J.}~\bibnamefont {Brand}},\ }\href {\doibase
  10.3390/condmat7010015} {\bibfield  {journal} {\bibinfo  {journal} {Condensed
  Matter}\ }\textbf {\bibinfo {volume} {7}} (\bibinfo {year} {2022}),\
  10.3390/condmat7010015}\BibitemShut {NoStop}%
\bibitem [{\citenamefont {Matveev}\ and\ \citenamefont
  {Furusaki}(2008)}]{matveev2008spectral}%
  \BibitemOpen
  \bibfield  {author} {\bibinfo {author} {\bibfnamefont {K.~A.}\ \bibnamefont
  {Matveev}}\ and\ \bibinfo {author} {\bibfnamefont {A.}~\bibnamefont
  {Furusaki}},\ }\href {\doibase 10.1103/PhysRevLett.101.170403} {\bibfield
  {journal} {\bibinfo  {journal} {Phys. Rev. Lett.}\ }\textbf {\bibinfo
  {volume} {101}},\ \bibinfo {pages} {170403} (\bibinfo {year}
  {2008})}\BibitemShut {NoStop}%
\bibitem [{\citenamefont {Reichert}\ \emph {et~al.}(2017)\citenamefont
  {Reichert}, \citenamefont {Petkovi\ifmmode~\acute{c}\else \'{c}\fi{}},\ and\
  \citenamefont {Ristivojevic}}]{PRBBose-Fermi}%
  \BibitemOpen
  \bibfield  {author} {\bibinfo {author} {\bibfnamefont {B.}~\bibnamefont
  {Reichert}}, \bibinfo {author} {\bibfnamefont {A.}~\bibnamefont
  {Petkovi\ifmmode~\acute{c}\else \'{c}\fi{}}}, \ and\ \bibinfo {author}
  {\bibfnamefont {Z.}~\bibnamefont {Ristivojevic}},\ }\href {\doibase
  10.1103/PhysRevB.95.045426} {\bibfield  {journal} {\bibinfo  {journal} {Phys.
  Rev. B}\ }\textbf {\bibinfo {volume} {95}},\ \bibinfo {pages} {045426}
  (\bibinfo {year} {2017})}\BibitemShut {NoStop}%
\bibitem [{\citenamefont {Schecter}\ \emph
  {et~al.}(2012{\natexlab{b}})\citenamefont {Schecter}, \citenamefont
  {Kamenev}, \citenamefont {Gangardt},\ and\ \citenamefont
  {Lamacraft}}]{PhysRevLett.108.207001}%
  \BibitemOpen
  \bibfield  {author} {\bibinfo {author} {\bibfnamefont {M.}~\bibnamefont
  {Schecter}}, \bibinfo {author} {\bibfnamefont {A.}~\bibnamefont {Kamenev}},
  \bibinfo {author} {\bibfnamefont {D.~M.}\ \bibnamefont {Gangardt}}, \ and\
  \bibinfo {author} {\bibfnamefont {A.}~\bibnamefont {Lamacraft}},\ }\href
  {\doibase 10.1103/PhysRevLett.108.207001} {\bibfield  {journal} {\bibinfo
  {journal} {Phys. Rev. Lett.}\ }\textbf {\bibinfo {volume} {108}},\ \bibinfo
  {pages} {207001} (\bibinfo {year} {2012}{\natexlab{b}})}\BibitemShut
  {NoStop}%
\bibitem [{\citenamefont {Zvonarev}\ \emph {et~al.}(2009)\citenamefont
  {Zvonarev}, \citenamefont {Cheianov},\ and\ \citenamefont
  {Giamarchi}}]{ZvonarevPhysRevB.80.201102}%
  \BibitemOpen
  \bibfield  {author} {\bibinfo {author} {\bibfnamefont {M.~B.}\ \bibnamefont
  {Zvonarev}}, \bibinfo {author} {\bibfnamefont {V.~V.}\ \bibnamefont
  {Cheianov}}, \ and\ \bibinfo {author} {\bibfnamefont {T.}~\bibnamefont
  {Giamarchi}},\ }\href {\doibase 10.1103/PhysRevB.80.201102} {\bibfield
  {journal} {\bibinfo  {journal} {Phys. Rev. B}\ }\textbf {\bibinfo {volume}
  {80}},\ \bibinfo {pages} {201102} (\bibinfo {year} {2009})}\BibitemShut
  {NoStop}%
\bibitem [{\citenamefont {Fuchs}\ \emph {et~al.}(2005)\citenamefont {Fuchs},
  \citenamefont {Gangardt}, \citenamefont {Keilmann},\ and\ \citenamefont
  {Shlyapnikov}}]{fuchs2005spin}%
  \BibitemOpen
  \bibfield  {author} {\bibinfo {author} {\bibfnamefont {J.~N.}\ \bibnamefont
  {Fuchs}}, \bibinfo {author} {\bibfnamefont {D.~M.}\ \bibnamefont {Gangardt}},
  \bibinfo {author} {\bibfnamefont {T.}~\bibnamefont {Keilmann}}, \ and\
  \bibinfo {author} {\bibfnamefont {G.~V.}\ \bibnamefont {Shlyapnikov}},\
  }\href {\doibase 10.1103/PhysRevLett.95.150402} {\bibfield  {journal}
  {\bibinfo  {journal} {Phys. Rev. Lett.}\ }\textbf {\bibinfo {volume} {95}},\
  \bibinfo {pages} {150402} (\bibinfo {year} {2005})}\BibitemShut {NoStop}%
\bibitem [{\citenamefont {Ristivojevic}(2022)}]{Zoran_YangG}%
  \BibitemOpen
  \bibfield  {author} {\bibinfo {author} {\bibfnamefont {Z.}~\bibnamefont
  {Ristivojevic}},\ }\href {\doibase 10.1103/PhysRevA.105.013327} {\bibfield
  {journal} {\bibinfo  {journal} {Phys. Rev. A}\ }\textbf {\bibinfo {volume}
  {105}},\ \bibinfo {pages} {013327} (\bibinfo {year} {2022})}\BibitemShut
  {NoStop}%
\bibitem [{\citenamefont {Trofimov}\ and\ \citenamefont
  {Peskov}(2009)}]{GPE_discretization}%
  \BibitemOpen
  \bibfield  {author} {\bibinfo {author} {\bibfnamefont {V.~A.}\ \bibnamefont
  {Trofimov}}\ and\ \bibinfo {author} {\bibfnamefont {N.~V.}\ \bibnamefont
  {Peskov}},\ }\href {\doibase 10.3846/1392-6292.2009.14.109-126} {\bibfield
  {journal} {\bibinfo  {journal} {Mathematical Modelling and Analysis}\
  }\textbf {\bibinfo {volume} {14}},\ \bibinfo {pages} {109} (\bibinfo {year}
  {2009})}\BibitemShut {NoStop}%
\bibitem [{\citenamefont {Patankar}(1980)}]{upwind_scheme}%
  \BibitemOpen
  \bibfield  {author} {\bibinfo {author} {\bibfnamefont {S.}~\bibnamefont
  {Patankar}},\ }\href {https://books.google.fr/books?id=5JMYZMX3OVcC} {\emph
  {\bibinfo {title} {Numerical Heat Transfer and Fluid Flow}}},\ Series in
  computational methods in mechanics and thermal sciences\ (\bibinfo
  {publisher} {Taylor \& Francis},\ \bibinfo {year} {1980})\BibitemShut
  {NoStop}%
\bibitem [{\citenamefont {Kamchatnov}\ \emph {et~al.}(2004)\citenamefont
  {Kamchatnov}, \citenamefont {Gammal},\ and\ \citenamefont
  {Kraenkel}}]{PhysRevA.69.063605}%
  \BibitemOpen
  \bibfield  {author} {\bibinfo {author} {\bibfnamefont {A.~M.}\ \bibnamefont
  {Kamchatnov}}, \bibinfo {author} {\bibfnamefont {A.}~\bibnamefont {Gammal}},
  \ and\ \bibinfo {author} {\bibfnamefont {R.~A.}\ \bibnamefont {Kraenkel}},\
  }\href {\doibase 10.1103/PhysRevA.69.063605} {\bibfield  {journal} {\bibinfo
  {journal} {Phys. Rev. A}\ }\textbf {\bibinfo {volume} {69}},\ \bibinfo
  {pages} {063605} (\bibinfo {year} {2004})}\BibitemShut {NoStop}%
\bibitem [{\citenamefont {Hoefer}\ \emph {et~al.}(2006)\citenamefont {Hoefer},
  \citenamefont {Ablowitz}, \citenamefont {Coddington}, \citenamefont
  {Cornell}, \citenamefont {Engels},\ and\ \citenamefont
  {Schweikhard}}]{PhysRevA.74.023623}%
  \BibitemOpen
  \bibfield  {author} {\bibinfo {author} {\bibfnamefont {M.~A.}\ \bibnamefont
  {Hoefer}}, \bibinfo {author} {\bibfnamefont {M.~J.}\ \bibnamefont
  {Ablowitz}}, \bibinfo {author} {\bibfnamefont {I.}~\bibnamefont
  {Coddington}}, \bibinfo {author} {\bibfnamefont {E.~A.}\ \bibnamefont
  {Cornell}}, \bibinfo {author} {\bibfnamefont {P.}~\bibnamefont {Engels}}, \
  and\ \bibinfo {author} {\bibfnamefont {V.}~\bibnamefont {Schweikhard}},\
  }\href {\doibase 10.1103/PhysRevA.74.023623} {\bibfield  {journal} {\bibinfo
  {journal} {Phys. Rev. A}\ }\textbf {\bibinfo {volume} {74}},\ \bibinfo
  {pages} {023623} (\bibinfo {year} {2006})}\BibitemShut {NoStop}%
\bibitem [{\citenamefont {Castro~Neto}\ and\ \citenamefont
  {Fisher}(1996)}]{castro_neto1996dynamics}%
  \BibitemOpen
  \bibfield  {author} {\bibinfo {author} {\bibfnamefont {A.~H.}\ \bibnamefont
  {Castro~Neto}}\ and\ \bibinfo {author} {\bibfnamefont {M.~P.~A.}\
  \bibnamefont {Fisher}},\ }\href {\doibase 10.1103/PhysRevB.53.9713}
  {\bibfield  {journal} {\bibinfo  {journal} {Phys. Rev. B}\ }\textbf {\bibinfo
  {volume} {53}},\ \bibinfo {pages} {9713} (\bibinfo {year}
  {1996})}\BibitemShut {NoStop}%
\bibitem [{\citenamefont {Petkovi\ifmmode~\acute{c}\else \'{c}\fi{}}\ and\
  \citenamefont {Ristivojevic}(2023)}]{MyPRLdissipative}%
  \BibitemOpen
  \bibfield  {author} {\bibinfo {author} {\bibfnamefont {A.}~\bibnamefont
  {Petkovi\ifmmode~\acute{c}\else \'{c}\fi{}}}\ and\ \bibinfo {author}
  {\bibfnamefont {Z.}~\bibnamefont {Ristivojevic}},\ }\href {\doibase
  10.1103/PhysRevLett.131.186001} {\bibfield  {journal} {\bibinfo  {journal}
  {Phys. Rev. Lett.}\ }\textbf {\bibinfo {volume} {131}},\ \bibinfo {pages}
  {186001} (\bibinfo {year} {2023})}\BibitemShut {NoStop}%
\bibitem [{\citenamefont {Ishikawa}\ and\ \citenamefont
  {Takayama}(1980)}]{ishikawa1980}%
  \BibitemOpen
  \bibfield  {author} {\bibinfo {author} {\bibfnamefont {M.}~\bibnamefont
  {Ishikawa}}\ and\ \bibinfo {author} {\bibfnamefont {H.}~\bibnamefont
  {Takayama}},\ }\href {\doibase 10.1143/JPSJ.49.1242} {\bibfield  {journal}
  {\bibinfo  {journal} {J. Phys. Soc. Jpn.}\ }\textbf {\bibinfo {volume}
  {49}},\ \bibinfo {pages} {1242} (\bibinfo {year} {1980})}\BibitemShut
  {NoStop}%
\end{thebibliography}

%

\end{document}